\documentclass[12pt]{article}

\global\arraycolsep=1pt

\usepackage{amsbsy,amssymb,latexsym,amsfonts,amsmath}
\usepackage{mathrsfs}
\usepackage{graphicx}
\usepackage{bm}

\numberwithin{equation}{section}
\newcommand{\bel}[1]{\begin{equation}\label{#1}}                     
\newcommand{\bal}[1]{\begin{eqnarray}\label{#1}}                     
\newcommand{\be}{\begin{equation}}
\newcommand{\ee}{\end{equation}}

\newcommand{\im}{\mathrm{i}}
\newcommand{\ex}{\mathrm{e}}
\newcommand{\de}{\mathrm{d}}
\newcommand{\dis}{\displaystyle}

\newcommand{\qq}{\qquad}

\renewcommand{\thefootnote}{\fnsymbol{footnote}}

\newcommand{\bea}{\begin{equation}}
\newcommand{\eea}{\end{equation}}

\def\a{\alpha}
\def\b{\beta}

\def\l{\lambda}

\begin{document}
%
%
\begin{titlepage}
\begin{flushright}
\normalsize
~~~~
OCU-PHYS 389\\
August, 2013 \\
\end{flushright}

\vspace{15pt}

\begin{center}
{\LARGE $2d$-$4d$ Connection between}\\
\vspace{5pt}
{\LARGE $q$-Virasoro/$W$ Block at Root of Unity Limit} \\
\vspace{5pt}  
{\LARGE   and     }\\
\vspace{5pt}  
{\LARGE Instanton Partition Function on ALE Space   }\\
\end{center}

\vspace{23pt}

\begin{center}
{ H. Itoyama$^{a, b}$\footnote{e-mail: itoyama@sci.osaka-cu.ac.jp},
T. Oota$^b$\footnote{e-mail: toota@sci.osaka-cu.ac.jp}
  and  R. Yoshioka$^b$\footnote{e-mail yoshioka@sci.osaka-cu.ac.jp}  }\\
%
\vspace{18pt}
%

$^a$ \it Department of Mathematics and Physics, Graduate School of Science\\
Osaka City University\\
\vspace{5pt}

$^b$ \it Osaka City University Advanced Mathematical Institute (OCAMI)

\vspace{5pt}

3-3-138, Sugimoto, Sumiyoshi-ku, Osaka, 558-8585, Japan \\

\end{center}
%
\vspace{20pt}
\begin{center}
Abstract\\
\end{center}
 We propose and demonstrate a limiting procedure in which, 
starting from  the $q$-lifted version 
(or $K$-theoretic five dimensional version) of the (W)AGT conjecture 
to be assumed in this paper, the Virasoro/$W$ block is generated 
in the $r$-th root of unity limit in $q$ in the $2d$ side, 
while the same limit automatically generates the projection
of the five dimensional instanton partition function onto that 
on the ALE space $\mathbb{R}^4/\mathbb{Z}_r$.
This circumvents case-by-case conjectures to be made 
in a wealth of examples found so far.
In the $2d$ side, we successfully generate the super-Virasoro algebra 
and the proper screening charge in the $q \rightarrow -1$, 
$t \rightarrow -1$ limit, from the defining relation 
of the $q$-Virasoro algebra and the $q$-deformed Heisenberg algebra.
The central charge obtained coincides with 
that of the minimal series carrying odd integers of 
the $N=1$ superconformal algebra. In the $r$-th root of unity limit 
in $q$ in the $2d$ side, we give some evidence of
the appearance of the parafermion-like currents.
Exploiting the $q$-analysis literatures, $q$-deformed $su(n)$ block is 
readily generated both at generic $q, t$ and the $r$-th root of unity limit.
In the $4d$ side, we derive the proper normalization function 
for general $(n, r)$ that accomplishes the automatic projection through
the limit.


\vfill

\setcounter{footnote}{0}
\renewcommand{\thefootnote}{\arabic{footnote}}

\end{titlepage}

\renewcommand{\thefootnote}{\arabic{footnote}}
\setcounter{footnote}{0}


\section{Introduction}
\label{sec:intro}


  Continuing attention has been paid to the correspondence between 
the two-dimensional conformal block \cite{BPZ1984} and 
the instanton sum \cite{Nak99,Nek0206} identified as 
the partition function of the four dimensional ${\cal N}=2$ 
supersymmetric gauge theory. 
  The both sides of this correspondence \cite{AGT0906,W0907} have 
already been intensively studied for more than a few years
and a wealth of such examples has been found by now. 
One of the central tools for our study is 
the $\beta$-deformed matrix model controlling 
the integral representation of the conformal block 
\cite{DV,IMO,MMS0911,MMS1001,IO5,MMM1003,IOYone,NR1112,FMP1210}
and the use \cite{IO5} of the formula \cite{kan96,kad97} 
on multiple integrals. 
   This general correspondence, on the other hand, 
has stayed as conjectures in most of the examples except
the few ones \cite{FL0912,HJS1004,MMS1012,MS1307} and 
one of the next steps in the developments would be to obtain
efficient understanding among these 
while avoiding making many conjectures.
  
 In this paper\footnote{Talks based on this work 
have already been given by the authors in the following workshops 
and the conference \cite{conference}.}, 
we regard the correspondence between
$q$-Virasoro \cite{SKAO,FR96,AKOS,AKMOS,AY,BP}/$W$ block 
versus five dimensional instanton partition function as a parent one.
  We propose the following procedure on the orbifolded examples 
of the correspondence 
\cite{BF1105,BMT,BBB1106,W1109,AT,I1110,BBFLT1111,BBT1211,ABT1306,bel0611}:
\begin{enumerate}
\item assume the  the $q$-lifted
 version (or $K$-theoretic five dimensional version) of the (W)AGT conjecture
 
\item
introduce the limiting procedure $q \rightarrow \omega$
 
\item
apply the same limiting procedure to $Z^{{\rm 5d}}_{{\rm inst}}$, which automatically generates the instanton partition function on ALE space.
\end{enumerate} 

  We emphasize that, through this limiting procedure (and the assumption (1)),
the resulting $2d$ conformal block is guaranteed to agree with the corresponding
instanton partition function on ALE space. 

\begin{figure}[h]
\begin{minipage}{10cm}
  \begin{center}
  \includegraphics[height=7cm]{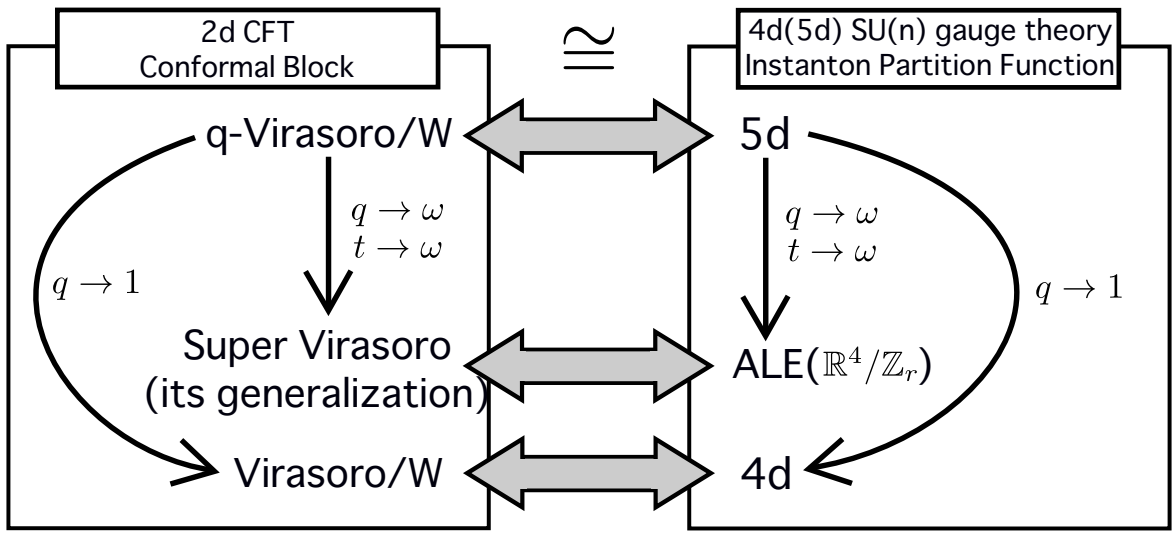} 
  \end{center}
\end{minipage}
\caption{Illustration of the procedure proposed in this paper.} 
\label{ComDiag}
\end{figure}

 The limiting procedure we propose in this paper is the following one (Eq. (\ref{limit:2d}) in the text):
\be
q = \omega \, \ex^{-(1/\sqrt{\beta}) h}, \qq
t = \omega \, \ex^{-\sqrt{\beta} h}, \qq
p = q/t = \ex^{Q_E h},
\ee
with $\omega = \ex^{2\pi \im/r}$. We introduce the root of unity limit $q \rightarrow \omega$
 by $h \rightarrow +0$ limit.
 Our procedure is illustrated in Figure \ref{ComDiag}.

In the next section, we first derive super-Virasoro algebra 
\cite{FQS,Eic,BKT},  starting from the defining relation of 
the $q$-Virasoro algebra and the $q$-deformed 
Heisenberg algebra \cite{SKAO}.
   The central charge obtained coincides with that of 
the minimal series \cite{FQS} carrying odd integers of 
the $N=1$ superconformal algebra. 

In section three, we consider the $r$-th root of unity limit of 
$q$ and $t$ for the $q$-Virasoro algebra, and we give some evidence of
the appearance of the parafermion-like currents.
 
 In section four, exploiting the $q$-analysis 
literatures \cite{War,FW,war0708,kan96,kad97},  
we introduce integral representation of 
$q$-deformed $su(n)$ block which is valid both at generic $q, t$ 
and the $r$-th root of unity limit.

In section five, we turn our attention to the $4d$ side. 
See, for instance, \cite{Nak94,Nak98,Nak9610,Nak9507,Nak9912,NY0306}.  
After some review on the ALE instanton partition function \cite{KN},  
we show that the same limit automatically generates the projection
of the five dimensional instanton partition function 
(see, for instance, \cite{AK}) onto that 
on the ALE space $\mathbb{R}^4/\mathbb{Z}_r$.
 We derive the proper normalization function for 
general $(n, r)$ that accomplishes 
the automatic projection through the limit.

For M theoretic view point of this correspondence, see,
for example, \cite{Tan1301}. For recent developments on the connection 
between $q$-deformed Virasoro algebra and the 5 dimensional partition
function on $S^5$, see \cite{NPP1303}.


\section{$\bm{q\rightarrow -1}$ limit of $\bm{q}$-Virasoro algebra and $\bm{N=1}$ SCFT}


In this section, we consider the $q \rightarrow -1$ limit of the $q$-Virasoro algebra
and discuss its connection with the $N=1$ super-Virasoro algebra.
This limit corresponds to the $r=2$ case of $q \rightarrow \omega = \ex^{2\pi \im/r}$.


\subsection{$\bm{q}$-Virasoro algebra for generic $\bm{q}$}


The $q$-Virasoro algebra $\mathrm{Vir}_{q,t}$ \cite{SKAO}
contains two parameters 
$q$ and $t$ and has a realization in terms of a $q$-deformed 
Heisenberg algebra $\mathcal{H}_{q,t}$:
\be
\begin{split}
[ \alpha_n, \alpha_m ] &= - \frac{1}{n} 
\frac{(1-q^n)(1-t^{-n})}{(1+p^n)} \delta_{n+m,0}, \qq (n \neq 0), \cr
[ \alpha_n, Q ] &= \delta_{n,0},
\end{split}
\ee
where $p=q/t$. In terms of these ``fundamental bosons'',
the generators of the $q$-Virasoro algebra $\mathcal{T}_n$ ($n \in \mathbb{Z}$)
are realized as
\bel{qVg}
\mathcal{T}(z) = : \exp\left( \sum_{ n \neq 0} \alpha_n z^{-n} \right):
p^{1/2} q^{\sqrt{\beta} \alpha_0}
+ : \exp\left( - \sum_{n \neq 0} \alpha_n (pz)^{-n} \right):
p^{-1/2} q^{-\sqrt{\beta} \alpha_0},
\ee
where $\mathcal{T}(z) = \sum_{n \in \mathbb{Z}} \mathcal{T}_n z^{-n}$,
and $\beta = \log t/\log q$.
Eq. \eqref{qVg} satisfies the defining relation
of the $q$-Virasoro algebra:
\be
f(z'/z) \mathcal{T}(z) \mathcal{T}(z') 
- f(z/z') \mathcal{T}(z') \mathcal{T}(z)
= \frac{(1-q)(1-t^{-1})}{(1-p)}
\Bigl[ \delta(pz/z') - \delta(p^{-1}z/z') \Bigr],
\ee
where
\be
f(z) = \exp\left( \sum_{n=1}^{\infty} \frac{1}{n}
\frac{(1-q^n)(1-t^{-n})}{(1+p^n)} z^n \right),
\ee
and the multiplicative delta function is defined by
\be
\delta(z) = \sum_{n \in \mathbb{Z}} z^n.
\ee
Let us introduce two $q$-deformed free bosons by
\bel{qfb}
\widetilde{\varphi}^{(\pm )}(z)
= \beta^{\pm 1/2}  \, Q + 2 \beta^{\pm 1/2}
\alpha_0 \log z + \sum_{n \neq 0} \frac{(1+p^{-n})}{(1-\xi_{\pm}{}^n)}
\alpha_{n} z^{-n},
\ee
where $\xi_+ = q$, $\xi_-=t$.

Their correlators are given by
\be
\begin{split}
\langle \widetilde{\varphi}^{(\pm)}(z_1)
\widetilde{\varphi}^{(\pm)}(z_2) \rangle
&= 2 \beta^{\pm 1} \log z_1 - \sum_{n=1}^{\infty}
\frac{(1+p^{\pm n})(1-\xi_{\mp}{}^n)}{n(1 - \xi_{\pm}{}^n)}
\left( \frac{z_2}{z_1} \right)^n \cr
&= \log\left[ z_1^{2\beta^{\pm 1}}
\left( 1 - \frac{z_2}{z_1} \right)
\frac{(p^{\pm 1} z_2/z_1; \xi_{\pm})_{\infty}}
{( \xi_{\mp} z_2/z_1; \xi_{\pm})_{\infty}} \right],
\end{split}
\ee
\be
\langle \widetilde{\varphi}^{(\pm)}(z_1)
\widetilde{\varphi}^{(\mp)}(z_2) \rangle
= \log(z_1-z_2) + \log(z_1 - p^{\mp 1} z_2).
\ee
Here
\be
(x; \xi)_{\infty} = \prod_{k=0}^{\infty} (1 - x \, \xi^k ).
\ee
Remark. By taking $q \rightarrow 1$ limit with $\beta$ fixed,
we have
\be
\lim_{q \rightarrow 1}
\langle \widetilde{\varphi}^{(\pm)}(z_1)
\widetilde{\varphi}^{(\pm)}(z_2) \rangle
= \log \Bigl[ (z_1 - z_2)^{2 \beta^{\pm 1}} \Bigr].
\ee

Two kinds of deformed screening currents are defined by
\bel{DSC}
S_{\pm}(z) = : \ex^{\pm \widetilde{\varphi}^{(\pm)}(z)} :.
\ee
The screening currents $S_{\pm}$
commute with the $q$-Virasoro generators
up to total $q$- or $t$-derivative:
\be
[ \mathcal{T}(z), S_+(z') ]
= -(1-q)(1-t^{-1})\frac{\de_q}{\de_q z'}
\Bigl[ \delta(z'/z) p^{-1/2} z' A_+(z') \Bigr],
\ee
\be
[ \mathcal{T}(z), S_-(z')]
= - (1-q^{-1})(1-t) \frac{\de_t}{\de_t z'}
\Bigl[ \delta(z'/z) p^{1/2} z' A_-(z') \Bigr],
\ee
where
\be
\begin{split}
A_+(z) &= : \exp\left( \sum_{n \neq 0}
\frac{(1+t^{n})}{(1-q^n)} \alpha_n z^{-n} \right):
\ex^{\sqrt{\beta} Q}
q^{-\sqrt{\beta} \alpha_0} z^{2 \sqrt{\beta} \alpha_0}, \cr
A_-(z) &= : \exp\left( - \sum_{n \neq 0}
\frac{(1+q^n)}{(1-t^{n})} \alpha_n (pz)^{-n} \right):
\ex^{-(1/\sqrt{\beta}) Q} q^{\sqrt{\beta} \alpha_0}
z^{- (2/\sqrt{\beta}) \alpha_0}.
\end{split}
\ee
We use the following convention of the $\xi$-derivative ($\xi=q,t$):
\be
\frac{\de_{\xi}}{\de_{\xi} z} f(z) = \frac{f(z) - f(\xi z)}{(1-\xi) z}.
\ee
Deformed screening charges \cite{AY} are defined by the Jackson integral
\be
Q^{+}_{[a,b]} = \int_a^b \de_q z\, S_+(z), \qq
Q^{-}_{[a,b]} = \int_a^b \de_t z\, S_-(z),
\ee
over a suitably chosen integral domain $[a,b]$.


\subsection{$\bm{q \rightarrow -1}$ limit of $\bm{q}$-bosons}


First, we consider the $q=-1$ limit of the $q$-bosons \eqref{qfb}.
Let us decompose them into ``even'' and odd parts:
\be
\widetilde{\varphi}^{(\pm)}(z) = \widetilde{\varphi}^{(\pm)}_{\mathrm{even}}(z)
+ \widetilde{\varphi}^{(\pm)}_{\mathrm{odd}}(z),
\ee
where
\be
\begin{split}
\widetilde{\varphi}_{\mathrm{even}}^{(\pm)}(z)&:=
\beta^{\pm 1/2} Q + \beta^{\pm 1/2}  \alpha_0 \log(z^2)
+ \sum_{n \neq 0} \frac{1+p^{-2n}}{1-\xi_{\pm}{}^{2n}} \alpha_{2n} z^{-2n}, \cr
\widetilde{\varphi}_{\mathrm{odd}}^{(\pm)}(z)&:=
\sum_{n \in \mathbb{Z}} \frac{1+p^{-2n-1}}{1-\xi_{\pm}{}^{2n+1}} \alpha_{2n+1}
z^{-2n-1}.
\end{split}
\ee
Note that
\be
\widetilde{\varphi}_{\mathrm{even}}^{(\pm)}(\ex^{\pi \im} z)
= \widetilde{\varphi}_{\mathrm{even}}^{(\pm)}(z) 
+ 2 \pi \im \, \beta^{\pm 1/2} \, 
\alpha_0, \qq
\widetilde{\varphi}_{\mathrm{odd}}^{(\pm)}(-z) = -
\widetilde{\varphi}_{\mathrm{odd}}^{(\pm)}(z).
\ee

Now we consider a $q\rightarrow -1$ limit. 
Simultaneously, $t \rightarrow -1$ limit is taken.
In the remaining part of this section, we assume that $\beta$
takes the following rational value:
\bel{ratbeta}
\beta = \frac{k_-+1/2}{k_++1/2} = \frac{2k_-+1}{2k_++1},
\ee
where $k_{\pm}$ are non-negative integers.
To make the limit definite, 
we choose the branch of logarithms of $q$ and $t$ as
\be
\log q = 2\pi \im ( k_+ + 1/2) - \frac{h}{\sqrt{\beta}}, 
\qq
\log t = \beta \log q = 2\pi \im ( k_- + 1/2) - \sqrt{\beta} h,
\ee
with $h > 0$. Also,
\be
\log p = \log q - \log t = 2\pi \im ( k_+ - k_-) + Q_E h.
\ee
Here $Q_E = \sqrt{\beta} - 1/\sqrt{\beta}$. Hence
\be
q = - \ex^{-(1/\sqrt{\beta})h}, \qq
t = - \ex^{-\sqrt{\beta} h}, \qq
p = q/t = \ex^{Q_E h}.
\ee
The  $q \rightarrow -1$ limit implies $h \rightarrow +0$ limit. 

We postulate the following limit
of the boson oscillators:
\be
\begin{split}
\alpha_{2n} &= - \frac{p^n}{|n|}
\sqrt{- \frac{(1-q^{2n})(1-t^{-2n})}{2(1+p^{2n})}} a_n = 
- h \, a_n + O(h^2), \qq
(n\neq 0), \cr
\alpha_0 &= a_0, \cr
\alpha_{2n+1} &= \frac{1}{n+1/2}
\sqrt{ \frac{(1-q^{2n+1})(1-t^{-(2n+1)})}{2(1+p^{-(2n+1)})}}\, 
\tilde{a}_{n+1/2} \cr
&=
\frac{1}{n+1/2} \, \tilde{a}_{n+1/2}  + Q_E \, h \, \tilde{a}_{n+1/2}
+ O(h^2).
\end{split}
\ee

This leads to the following standard commutation relations
of boson and twisted boson oscillators:
\be
[ a_n, a_m] = n \delta_{n+m,0}, \qq [ a_n, Q ] = \delta_{n,0},
\ee
\bel{TBO}
[ \tilde{a}_{n+1/2}, \tilde{a}_{-m-1/2} ] = ( n + 1/2) \delta_{n,m}.
\ee
Therefore, in the $q\rightarrow -1$ limit of $q$-bosons \eqref{qfb}, 
we obtain two free bosons
$\phi(w)$ and $\varphi(w)$:
\bel{L2qB}
\begin{split}
\widetilde{\varphi}_{\mathrm{even}}^{(\pm)}(z)
&= \beta^{\pm 1/2} \phi(w)
+ O(h), \cr
\widetilde{\varphi}_{\mathrm{odd}}^{(\pm)}(z)
&= \varphi(w) + O(h),
\end{split}
\ee
where $z=\sqrt{w}$, and
\be
\phi(w) = Q + a_0 \log w - \sum_{n \neq 0}  \frac{a_n}{n} w^{-n},
\ee
\bel{varphi}
\varphi(w) 
= \sum_{n \in \mathbb{Z}} \frac{\tilde{a}_{n+1/2}}{n+1/2}
w^{-n-1/2}.
\ee
Here $\phi(w)$ (resp. $\varphi(w)$) 
is the ordinary (resp. twisted) boson on the $w$-plane:
\be
\langle \phi(w_1) \phi(w_2) \rangle = \log(w_1-w_2),
\qq
\langle \varphi(w_1) \varphi(w_2) \rangle = \log\left(
\frac{\sqrt{w_1}-\sqrt{w_2}}{\sqrt{w_1}+\sqrt{w_2}} \right).
\ee
Note that $\phi(\ex^{2\pi \im} w) = \phi(w)+2\pi \im\, a_0$, 
$\varphi(\ex^{2\pi \im } w) = - \varphi(w)$.


\subsection{$\bm{q \rightarrow -1}$ limit of screening currents}


Next, we consider the $q \rightarrow -1$ limit of the screening
currents \eqref{DSC}. Using the limit of $q$-bosons \eqref{L2qB},
we easily see that
\be
\lim_{q \rightarrow -1} S_+(z)
= : \ex^{\sqrt{\beta} \phi(w)} \ex^{\varphi(w)}:, \qq
\lim_{q \rightarrow -1} S_-(z)
= : \ex^{-(1/\sqrt{\beta}) \phi(w)} \ex^{-\varphi(w)} :,
\ee
where $z=\sqrt{w}$. 

First, we rewrite the vertex operators $: \ex^{\pm \varphi(w)}:$.
Since the weight $\pm 1$ is in a (one-dimensional) integral
lattice $(\pm 1)^2 = 1 \in \mathbb{Z}$, 
we can construct two fermions\footnote{If the weights of the vertex
operators were
in the even lattice, $(\pm \sqrt{2})^2 = 2 \in 2 \mathbb{Z}$, 
one could construct the (level $1$) 
basic representation of the affine $SU(2)$ algebra
by using $\partial \varphi(w)$ and $: \ex^{\pm \sqrt{2} \varphi(w)}:$ 
\cite{LW78}.}
:
\bel{defNSR}
\psi(w)\equiv \frac{\im}{2 \sqrt{2 w}}
\Bigl( : \ex^{\varphi(w)}: - : \ex^{-\varphi(w)} : \Bigr),
\qq
\widehat{\psi}(w)\equiv 
\frac{1}{2\sqrt{2 w}}\Bigl( : \ex^{\varphi(w)} :
+ :\ex^{-\varphi(w)}:\Bigr).
\ee
In terms of these fermions, we have the twisted boson/fermion correspondence:
\be
: \ex^{ \pm \varphi(w)}: = \sqrt{2w} \Bigl( \widehat{\psi}(w)
 \mp \im \psi(w) \Bigr).
\ee

Using the commutation relations \eqref{TBO}, we can show that
these two fermions \eqref{defNSR}
obey the following anti-commutation relations:
\bel{NSR}
\begin{split}
\{ \psi(w_1), \psi(w_2) \} &= \frac{1}{w_1} \delta(w_2/w_1), \cr
\{ \widehat{\psi}(w_1), \widehat{\psi}(w_2) \}
&= \frac{1}{\sqrt{w_1 w_2}} \delta(w_2/w_1), \cr
\{ \psi(w_1), \widehat{\psi}(w_2) \} &= 0.
\end{split}
\ee

Notice that
\be
\psi(\ex^{2\pi \im} \, w) = \psi(w), \qq
\widehat{\psi}(\ex^{2\pi \im} \, w) = - \widehat{\psi}(w).
\ee
They admit the following mode expansion on the $w$-plane:
\be
\psi(w) = \sum_{r \in \mathbb{Z}+1/2} \psi_r \, w^{-r-1/2}, \qq
\widehat{\psi}(w) = \sum_{n \in \mathbb{Z}}
\widehat{\psi}_n \, w^{-n-1/2}.
\ee
In terms of these modes, the anti-commutation relations \eqref{NSR}
are rewritten as
\be
\{ \psi_r, \psi_s \} = \delta_{r+s,0}, \qq
\{ \widehat{\psi}_m, \widehat{\psi}_n \} = \delta_{m+n,0}, \qq
\{ \psi_r, \widehat{\psi}_m \} = 0.
\ee
Here $r, s \in \mathbb{Z}+1/2$ and $m,n \in \mathbb{Z}$.
Hence, $\psi(w)$ (resp. $\widehat{\psi}(w)$) is an NS fermion
(resp. an R fermion) on the $w$-plane.

On the Fock vacuum,
\be
\bigl\langle 0 | \psi(w_1) \psi(w_2) | 0 \bigr\rangle
= \frac{1}{w_1-w_2}, \qq
\bigl\langle 0| \widehat{\psi}(w_1) \widehat{\psi}(w_2) 
|0 \bigr\rangle
= \frac{1}{2(w_1-w_2)}\left( \sqrt{\frac{w_1}{w_2}}
+ \sqrt{\frac{w_2}{w_1}} \right), 
\ee
and $\bigl\langle 0| \psi(w_1) \widehat{\psi}(w_2) |0 \bigr\rangle = 0$.

\vspace{5mm}

\noindent
Remark. Let $|0'\rangle= \sqrt{2} \, \tilde{a}_{-1/2} | 0 \rangle$.
By definition \eqref{defNSR}, $\psi_r$ (resp. $\widehat{\psi}_n$) is
constructed as an infinite sum of terms with odd (resp. even) number of 
twisted boson oscillators  $\{ \tilde{a}_{n+1/2}\}$.
Hence, the state $|0' \rangle$ is not contained in 
the ``Ramond'' sub-module $\mathcal{F}_R$ of the twisted boson module
over the Fock vacuum:
\be
|0' \rangle \notin
\mathcal{F}_R = \mathbb{C}[\widehat{\psi}_{-1}, \widehat{\psi}_{-2},
\dotsm ] | 0 \rangle,
\qq \mathcal{F}_R  \subset \mathbb{C}[ \tilde{a}_{-1/2},
\tilde{a}_{-3/2}, \dotsc] | 0 \rangle.
\ee
But the state $|0 ' \rangle$ can be obtained by the action of
the NS fermion: $\psi_{-1/2} | 0 \rangle = - \im \, | 0 ' \rangle$. 
Moreover, we have the following isomorphism 
between twisted boson module and NS and R fermion module:
\be
\mathbb{C}[ \psi_{-1/2}, \psi_{-3/2}, \dotsm, \widehat{\psi}_{-1},
\widehat{\psi}_{-2}, \dotsm ] | 0 \rangle
= \mathbb{C}[ \tilde{a}_{-1/2}, \tilde{a}_{-3/2}, \dotsm] | 0 \rangle.
\ee
One can show that
\be
\widehat{\psi}_0 | 0 \rangle = \frac{1}{\sqrt{2}} | 0 \rangle, \qq
\widehat{\psi}_0 | 0' \rangle = - \frac{1}{\sqrt{2}} | 0' \rangle,
\ee
\be
\langle 0' | \widehat{\psi}(w_1) \widehat{\psi}(w_2) 
| 0' \rangle
= \frac{1}{2(w_1-w_2)}\left( \sqrt{\frac{w_1}{w_2}}
+ \sqrt{\frac{w_2}{w_1}} \right),
\ee
where $\langle 0' | = \sqrt{2}\, \langle 0 | \tilde{a}_{1/2}$.

\subsection{$\bm{q=-1}$ limit of $\bm{q}$-Virasoro generators}

In this subsection, we consider the $q \rightarrow -1$ limit
of the generating function $\mathcal{T}(z)$
of the $q$-Virasoro generators
\eqref{qVg}. It behaves in this limit as
\be
\begin{split}
& (-1)^{k_+-k_-} \mathcal{T}(z) \cr 
&=\, \, 
\sqrt{2} \, \ex^{2\pi \im \sqrt{\beta}(k_++1/2) a_0}
 \Bigl[ w^{1/2} \bigl( \widehat{\psi}(w) - \im \psi(w) \bigr)
- h \, w^{3/2} \Bigl( \widehat{G}(w) - \im \, G(w) \Bigr) \Bigr] \cr
& + \sqrt{2}\, \ex^{- 2\pi \im \sqrt{\beta}(k_++1/2) a_0}
\Bigl[ w^{1/2} \bigl( \widehat{\psi}(w) + \im \psi(w) \bigr)
+ h \, w^{3/2} \Bigl( \widehat{G}(w) + \im \, G(w) \Bigr) \Bigr] + O(h^2)
\cr
&=
2 \sqrt{2}\, \cos \bigl( 2\pi \sqrt{\beta}(k_++1/2) a_0\bigr)
\Bigl( w^{1/2} \widehat{\psi}(w) +\im \, h\, w^{3/2}\,  G(w) \Bigr) \cr
&\  + 2 \sqrt{2}\, \sin \bigl( 2\pi \sqrt{\beta}(k_++1/2) a_0\bigr)
\Bigl( w^{1/2} \psi(w) - \im \, h\, w^{3/2} \, \widehat{G}(w) \Bigr) 
+ O(h^2),
\end{split}
\ee
where 
\be
\begin{split}
G(w)&= \psi(w) \partial \phi(w) + Q_E \, \partial \psi(w), \cr
\widehat{G}(w) 
&= \widehat{\psi}(w) \partial \phi(w) + Q_E \, \partial
\widehat{\psi}(w).
\end{split}
\ee

Let
\bel{Tw}
T(w) = \frac{1}{2} : \bigl( \partial \phi(w) \bigr)^2:
+ \frac{Q_E}{2} \partial^2 \phi(w) 
- \frac{1}{2} :\psi(w) \partial \psi(w):.
\ee
Then, $T(w)$ and $G(w)$ 
obey the $N=1$ superconformal algebra in 
the NS sector:
\be
\begin{split}
T(w_1) T(w_2) &= \frac{(3/4) \hat{c}}{(w_1-w_2)^4} 
+ \frac{2\, T(w_2)}{(w_1-w_2)^2}
+ \frac{T'(w_2)}{w_1-w_2} + \dotsm, \cr
T(w_1) G(w_2) &= \frac{(3/2) G(w_2)}{(w_1-w_2)^2}
+ \frac{G'(w_2)}{w_1-w_2} + \dotsm, \cr
G(w_1) G(w_2) &= \frac{\hat{c}}{(w_1-w_2)^3} + \frac{2T(w_2)}{w_1-w_2}+
\dotsm,
\end{split}
\ee
where
\bel{scc}
\hat{c} = 1 -2 Q_E^2 
= 1 -2 \left( \sqrt{\beta} -
\frac{1}{\sqrt{\beta}}\right)^2
= 1 - \frac{8( k_- - k_+)^2}{(2k_++1)(2k_-+1)}.
\ee
After replacing $\psi(w)$ with $\widehat{\psi}(w)$ in \eqref{Tw}, 
$T(w)$ and $\widehat{G}(w)$ obey the $N=1$ algebra in the R sector.

Recall that
the $N=1$ minimal superconformal models have the central charge
\be
\hat{c} = 1 - \frac{2(m'-m)^2}{mm'}.
\ee
Therefore, $\hat{c}$ \eqref{scc} corresponds to an $N=1$ minimal
model such that $m$ and $m'$ are both
positive odd integers: $m=2k_++1$ and $m'=2k_-+1$. 
Without loss of generality, we can take $m'>m$, i.e., $k_->k_+$.
The unitary minimal models \cite{FQS} occur when $k_-=k_++1$ and
$k_+ \geq 1$.


\subsection{Vertex operator and its $\bm{q \rightarrow -1}$ limit}


Let us choose a $q$-deformed Vertex operator $V_{\alpha}(z)$
as
\bel{qVOvl}
V_{\alpha}(z) = : \ex^{\Phi_{\alpha}(z)}:,
\ee
where
\be
\Phi_{\alpha}(z) = \alpha \, Q + 2 \alpha \, \alpha_0 \log z
+ \sum_{n \neq 0}
\frac{q^{-n}(1-q^{2\sqrt{\beta} \alpha |n|})}
{(1-q^{-|n|})(1-t^{-n})} \alpha_n z^{-n}.
\ee
The correlator of this $q$-boson with
$\widetilde{\varphi}^{(+)}$ is given by
\be
\begin{split}
\langle \Phi_{\alpha}(z_1) \widetilde{\varphi}^{(+)}(z_2) \rangle
= \langle \widetilde{\varphi}^{(+)}(z_1) \Phi_{\alpha}(z_2) \rangle
&= 2 \sqrt{\beta} \, \alpha \log z_1 - \sum_{n=1}^{\infty}
\frac{1}{n} \frac{(1-q^{2\sqrt{\beta} \alpha n})}{(1-q^n)}
\left( \frac{qz_2}{z_1} \right)^n  \cr
&= \log \left[ z^{2\sqrt{\beta} \alpha}
\frac{(qz_2/z_1; q)_{\infty}}{(q^{1+2\sqrt{\beta} \alpha} z_2/z_1;
  q)_{\infty}}
\right].
\end{split}
\ee
The correlator among themselves is given by
\be
\langle \Phi_{\alpha}(z_1) \Phi_{\alpha'}(z_2) \rangle
= 2 \alpha \alpha' \log z_1 - \sum_{n=1}^{\infty}
\frac{1}{n}\frac{(1-q^{2\sqrt{\beta} \alpha n})
(1-q^{2\sqrt{\beta} \alpha' n})}
{(1-q^n)(1-t^n)(1+p^n)} \left( \frac{q^2 z_2}{z_1} \right)^n.
\ee

We restrict the parameter $\alpha$ to take values corresponding to those of 
the primary fields of the minimal theories in the NS sector: 
\be
\alpha = \alpha_{r,s} = - \left( \frac{1-r}{2} \right)
\frac{1}{\sqrt{\beta}}
+ \left( \frac{1-s}{2} \right) \sqrt{\beta},
\ee
where
\be
1 \leq r \leq 2k_-, \qq
1 \leq s \leq 2k_+, \qq
r-s \in 2 \mathbb{Z}.
\ee
Then
\be
L_{r,s} \equiv (2k_++1)\sqrt{\beta} \alpha_{r,s}
= - k_+(1-r) + (1-s) k_- + \left( \frac{r-s}{2} \right) \in
\mathbb{Z}.
\ee
We can see that
\be
\lim_{q \rightarrow -1} \Phi_{\alpha_{r,s}}(z)
= \alpha_{r,s} \phi(w) - \frac{1}{4} \Bigl(1- (-1)^{L_{r,s}} \Bigr)
\varphi(w).
\ee
Therefore, the $q \rightarrow -1$ limit of the
deformed vertex operator \eqref{qVOvl} is given by
\be
\lim_{q \rightarrow -1}
V_{\alpha_{r,s}}(z)
= 
\begin{cases}
: \ex^{\alpha_{r,s} \phi(w)}: & \ \ \ \mbox{for $L_{r,s}$ even}, \cr
: \ex^{\alpha_{r,s} \phi(w)} \ex^{-(1/2) \varphi(w)}: & \ \ \ \mbox{for $L_{r,s}$
    odd}.
\end{cases}
\ee
For $L_{r,s}$ even, $: \ex^{\alpha_{r,s}(w)}:$ is 
exactly equal to the Coulomb gas representation of
the bosonic primary field in the NS sector with scaling dimension
\be
\Delta_{\alpha_{r,s}} = \frac{1}{2} \alpha_{r,s}( \alpha_{r,s} -
Q_E)
= - \frac{1}{8} Q_E^2 + \frac{1}{8} \left(
- \frac{r}{\sqrt{\beta}} + s \sqrt{\beta} \right)^2.
\ee

For $L_{r,s}$ odd, the interpretation of the vertex operator
$:\ex^{-(1/2) \varphi(w)}:$ is unclear. The operator product expansion
of the vertex operators
$:\ex^{\pm (1/2) \varphi(w)}:$ with the NS fermion is given by
\be
\psi(w_1) :\ex^{\pm (1/2) \varphi(w_2)}:
= \mp \frac{\im}{\sqrt{2}(w_1-w_2)^{1/2}} : \ex^{\mp (1/2)
  \varphi(w_2)}:
+ \dotsm,
\ee
which is similar to those of the spin fields, 
but they do not affect the Fock vacuum:
\be
: \ex^{\pm (1/2) \varphi(0)}: | 0 \rangle = | 0 \rangle.
\ee
Hence, these are not spin fields. 
In the remaining part of this section, we 
only consider the case of $L_{r,s}$ even.

Next, let us consider the $q \rightarrow -1$ limit of the Jackson integral
\be
\int_0^a \de_q z\, f(z) = a(1-q) \sum_{k=0}^{\infty} f(aq^k) q^k.
\ee
For $q=-\ex^{-h}$ and $f(z) = \sum_{n=0}^{\infty} f_n z^n$,
we can show that
\be
\lim_{h \rightarrow 0} (1+q) \int_0^a \de_q z \, f(z)
= \int_0^a \bigl( f(z) - f(-z) \bigr) \de z
= \int_0^a \left( \frac{f(z) - f(-z)}{z} \right) z \de z
= \int_0^{a^2} g(w) \de w.
\ee
Here $w=z^2$ and
\be
g(w):= \frac{f(z) - f(-z)}{2z}.
\ee

We propose the following ``limit'' of the
$q$-screening charge $Q_{[a,b]}^+$
for the rational $\beta$ \eqref{ratbeta}:
\bel{Qproj}
\begin{split}
& \frac{\im}{\sqrt{2}} (1+q) Q^{+}_{[a,b]}
= \frac{\im}{\sqrt{2}} (1+q) \int_a^b \de_q z\,
: \ex^{\widetilde{\varphi}_{\mathrm{odd}}(z)}:
\, : \ex^{\widetilde{\varphi}_{\mathrm{even}}(z)}: \cr
&\longrightarrow
\frac{\im}{\sqrt{2}} \int_a^b \de z \, z 
\left(\frac{: \ex^{\varphi(z)}: - :\ex^{-\varphi(z)}:}{z} \right)
: \ex^{\sqrt{\beta} \phi(z^2)}:
\equiv Q_{[a^2,b^2]}^{(+)},
\end{split}
\ee
where
\bel{SSC}
Q_{[a^2,b^2]}^{(+)}
= \int_{a^2}^{b^2} \de w \, \psi(w) : \ex^{\sqrt{\beta} \phi(w)} :.
\ee
Here we have ignored the subtlety due to the
zero-mode and negative power terms. To be more precise,
the ``limit'' \eqref{Qproj} includes
a kind of projection
imposed by hand.  Note that
\be
\lim_{h \rightarrow 0} : \ex^{\widetilde{\varphi}_{\mathrm{odd}}(z)}:
= : \ex^{\varphi(z)}:
= \sqrt{2} z \left( \widehat{\psi}(w) - \im \psi(w) \right).
\ee
Hence this projection is equivalent to drop the R fermion
$\widehat{\psi}(w)$.
We see that the resulting
screening charge \eqref{SSC} is the screening charge
for the superconformal block \cite{KIKKMO,AGZ}.


\section{Root of unity limit of $\bm{q}$-Virasoro algebra and parafermion-like structure}


In this section, we consider the $r$-th root of unity limit of the
$q$-Virasoro algebra. In particular, we show that the screening currents
$S_{\pm}$ in the limit are closely related to parafermion-like currents.

The parafermion currents are introduced in \cite{ZF85}.
The partition functions of the parafermionic theories are studied in
\cite{GQ87}. For recent work which discuss the connection between
parafermions and Selberg integrals, see \cite{BFL1011}.

We assume that $\beta$ takes the following rational value:
\be
\beta = \frac{k_- + 1/r}{k_+ + 1/r} = \frac{r k_- +1}{r k_+ + 1},
\ee
where $k_{\pm}$ are non-negative integers and choose the branch of logarithms
of $q$ and $t$ as
\be
\log q = 2 \pi \im (k_+ + 1/r) - \frac{h}{\sqrt{\beta}}, \qq
\log t = \beta \log q = 2\pi \im ( k_- + 1/r) - \sqrt{\beta} h,
\ee
with $h>0$. Hence
\be
q = \ex^{2\pi \im/r} \, \ex^{-(1/\sqrt{\beta}) h}, \qq
t = \ex^{2\pi \im/r}  \, \ex^{-\sqrt{\beta} h}, \qq
p = q/t = \ex^{Q_E h}.
\label{limit:2d}
\ee
The root of unity limit $q \rightarrow \ex^{2\pi \im/r}$ implies $h \rightarrow +0$ limit.
The $q \rightarrow -1$ limit in the previous section corresponds to the $r=2$
case.


\subsection{$q\rightarrow \ex^{2\pi \im/r}$ limit of $q$-bosons}


Let us decompose the $q$-deformed free bosons \eqref{qfb} into two parts:
\be
\widetilde{\varphi}^{(\pm)}(z) = \widetilde{\varphi}^{(\pm)}_0(z)
+ \widetilde{\varphi}^{(\pm)}_R(z),
\ee
where
\be
\begin{split}
\widetilde{\varphi}^{(\pm)}_0(z) &= \beta^{\pm 1/2} Q
+ \frac{2}{r} \beta^{\pm 1/2} \alpha_0 \log z^r
+ \sum_{n \neq 0} \frac{(1+p^{-nr})}{(1-\xi_{\pm}^{nr})} \alpha_{nr} \, z^{-nr},
\cr
\widetilde{\varphi}^{(\pm)}_R(z)
&= \sum_{\ell=1}^{r-1} \sum_{n \in \mathbb{Z}}
\frac{(1+p^{-nr-\ell})}{1-\xi_{\pm}^{nr+\ell}} \alpha_{nr+\ell} \, z^{-nr-\ell}.
\end{split}
\ee

We rescale the oscillators as follows
\be
\begin{split}
Q &= \sqrt{\frac{2}{r}} \, Q_0, \cr
\alpha_0 &= \sqrt{\frac{r}{2}} \, a_0, \cr
\alpha_{nr} &= - \sqrt{\frac{r}{2}} \, h \, a_n + O(h^2), \qq (n\neq 0)\cr
\alpha_{nr+\ell} &= \sqrt{\frac{2}{r}}
\frac{(1-\ex^{2\pi \im(nr+\ell)/r})}{2(n+\ell/r)}
\tilde{a}_{n+\ell/r} + O(h), \qq \qq (\ell=1,2,\dotsc, r-1).
\end{split}
\ee
The rescaled oscillators obey the following commutation relations
\be
[ a_m, a_n ] = m \delta_{m+n,0}, \qq
[ a_n, Q_0 ] = \delta_{n,0},
\ee
\be
[ \tilde{a}_{n+\ell/r}, \tilde{a}_{-m - \ell'/r} ]
= (n + \ell/r) \delta_{m,n} \delta_{\ell, \ell'}.
\ee

In the $q \rightarrow \ex^{2\pi \im/r}$ limit, we have
\be
\begin{split}
\lim_{h \rightarrow +0} \widetilde{\varphi}_0^{(\pm)}(z) &= \sqrt{\frac{2}{r}}\, 
\beta^{\pm 1/2} \phi(w), \cr
\lim_{h \rightarrow +0} \widetilde{\varphi}^{(\pm)}_R(z)
&= \sqrt{\frac{2}{r}} \, \varphi(w),
\end{split}
\ee
where $w=z^r$ and 
\be
\phi(w) = Q_0 + a_0 \log w  - \sum_{n \neq 0} \frac{a_n}{n} w^{-n},
\ee
\be
\varphi(w) = \sum_{\ell=1}^{r-1} \varphi^{(\ell)}(w),
\qq
\varphi^{(\ell)}(w) = \sum_{n \in \mathbb{Z}}
\frac{\tilde{a}_{n+\ell/r}}{n+\ell/r} w^{-n-\ell/r}.
\ee
The correlation functions are given by
\be
\langle \phi(w_1) \phi(w_2) \rangle = \log(w_1-w_2),
\ee
\be
\langle \varphi^{(\ell)}(w_1) \varphi^{(\ell')}(w_2) \rangle
= \delta_{\ell+\ell',r} \sum_{k=0}^{r-1}
\omega^{-k \ell} \log
\left[ 1 - \omega^k \left( \frac{w_2}{w_1} \right)^{1/r} \right].
\ee

\be
\langle \varphi(w_1) \varphi(w_2) \rangle
= \log\left[ \frac{(1-(w_2/w_1)^{1/r})^r}{1-(w_2/w_1)} \right]
= \log\left[ \frac{(w_1^{1/r}-w_2^{1/r})^r}{w_1-w_2} \right].
\ee


\subsection{$q\rightarrow \ex^{2\pi \im/r}$ limit of screening currents}


In the $q \rightarrow \ex^{2\pi \im/r}$ limit, the screening currents \eqref{DSC}
turn into
\be
\lim_{q \rightarrow \ex^{2\pi \im/r}}
S_{\pm}(z) = : \exp\left( \pm \sqrt{\frac{2}{r}} \beta^{\pm 1/2} \phi(w) \right):\, 
: \exp\left( \pm \sqrt{\frac{2}{r}} \varphi(w) \right) :.
\ee

Using the vertex operators $\ex^{\pm \sqrt{2/r} \, \varphi(w)}$, we can introduce the following fields
\be
\Psi_{\pm}^{(\ell)}(w):= \frac{1}{r^{3/2} w^{1-(1/r)}}
\sum_{k=0}^{r-1} \ex^{2\pi \im k\ell/r}
:\exp\left( \pm \sqrt{\frac{2}{r}} \varphi( \ex^{2\pi \im k} w ) \right) :.
\ee
Note that $\Psi_{\pm}^{(\ell+r)} = \Psi_{\pm}^{(\ell)}$. Hence we can choose the
range of $\ell$ as $\ell=0,1, \dots, r-1$.

These fields have the following periodicity:
\be
\Psi^{(\ell)}_{\pm}(\ex^{2\pi \im} w)
= \ex^{- 2\pi \im (\ell - r+1)/r} \Psi^{(\ell)}_{\pm}(w).
\ee

The correlators for these fields are given by
\be
\begin{split}
& \langle 0 | \Psi_{\pm}^{(\ell_1)}(w_1)
\Psi_{\mp}^{(\ell_2)}(w_2) | 0 \rangle \cr
&= \frac{\delta_{\ell_1+\ell_2,r}}{(w_1-w_2)^{2(1-(1/r))}}
\left(\frac{w_2}{w_1} \right)^{(\ell_1+1-r)/r}
\left[ 1 + \left( \frac{\ell_1+1-r}{r} \right) \frac{(w_1-w_2)}{w_1} \right],
\end{split}
\ee
\be
\begin{split}
& \langle 0 | \Psi_{\pm}^{(\ell_1)}(w_1)
\Psi_{\pm}^{(\ell_2)}(w_2) | 0 \rangle \cr
&=\frac{1}{r (w_1-w_2)^{2/r} (w_1 w_2)^{1-(1/r)} }
\left[ \delta_{\ell_1,0} \delta_{\ell_2,0}
w_1^{2/r} - 2 \delta_{\ell_1,1} \delta_{\ell_2,r-1}
(w_1 w_2)^{1/r}
+ \delta_{\ell_1,2} \delta_{\ell_2, r-2} w_2^{2/r} \right].
\end{split}
\ee
 
Note that for $w_1 \rightarrow w_2$,
\be
\langle 0 | \Psi_{\pm}^{(\ell_1)}(w_1)
\Psi_{\mp}^{(\ell_2)}(w_2) | 0 \rangle
= \frac{\delta_{\ell_1+\ell_2,r}}
{ (w_1-w_2)^{2(1-(1/r))}}
\left\{ 1 + O \Bigl( (w_1-w_2)^2 \Bigr) \right\}.
\ee
These fields $\Psi^{(\ell)}_{\pm}(w)$ are analogue of the parafermion current
with scaling dimension $\Delta_1 = 1-(1/r)$.
 
For example, if we set
\bel{paraF}
\Psi(w):= \Psi_+^{(1)}(w), \qq
\overline{\Psi}(w):= \Psi_-^{(r-1)}(w),
\ee
they obey (for $r>3$)
\be
\langle 0| \overline{\Psi}(w_1) \Psi(w_2) | 0 \rangle
= \frac{1}{(w_1-w_2)^{2(1-(1/r))}}, 
\ee
\be
\langle 0 | \Psi(w_1) \Psi(w_2) | 0 \rangle = 0, \qq
\langle 0 | \overline{\Psi}(w_1) \overline{\Psi}(w_2) | 0 \rangle
= 0.
\ee

For $f(z) = \sum_{n=0}^{\infty} f_n z^n$, the Jackson integral
in the $q \rightarrow \ex^{2\pi \im/r}$ limit is given by
\be
\lim_{h \rightarrow 0} \frac{(1-q^r)}{(1-q)}
\int_0^a \de_q z \, f(z) = \int_0^{a^r} \de w \, g(w),
\ee
where
\be
g(w) = \frac{1}{rz^{r-1}} \sum_{k=0}^{\infty} \ex^{2\pi\im k/r}
f(\ex^{2\pi \im k/r} z) = \sum_{n=0}^{\infty} f_{nr+r-1} w^n.
\ee
Inspired by this limit of the Jackson integral,  
it may be useful to consider the following ``projection"
\be
\frac{1}{\sqrt{r}} \frac{(1-q^r)}{(1-q)}\int_a^b \de_q z \, S_+(z) 
\longrightarrow Q_{[a^r,b^r]}^{(+)} =\int_{a^r}^{b^r} \de w\,
\Psi(w) \, : \ex^{\sqrt{2\beta/r}\,  \phi(w)}:, 
\ee
which is a generalization of \eqref{Qproj}.

For the $r=2$ case, $Q_{[a^2,b^2]}^{(+)}$ is the screening charge of the
$N=1$ superconformal algebra. Hence $Q_{[a^r, b^r]}^{(+)}$ may play the role
of a screening charge for a parafermionic algebra.


\section{Review of conformal block and its $\bm{q}$-lift}


The $q$-deformed $W_n$ algebra is introduced in \cite{FF95,AKOS}.
The $q$-Virasoro algebra is the $n=2$ case of these series of algebras.
For general $n$, the root of unity limit can be studied
similarly as in sections 2 and 3. But the analysis of the limit requires a hard
task. 
 
The $q$-deformed $W$ algebra at roots of unity itself will not be exploited here
for our study of the $q$-lifted version of (W)AGT conjecture.
It is the ($q$-deformed) conformal block for the $W_n$ algebra which play
the key role for the conjecture. 

For the conformal block, there is a simple recipe for the $q$-lift ($q$-deformation)
without explicitly treating the $q$-$W$ algebra.
In subsection 4.1, we review the conformal block of $W_n$ algebra
written as the Dotsenko-Fateev (DF) multiple integrals.
It can be converted into the multiple integrals closely related to  
the $A_{n-1}$ Selberg integrals.
In subsection 4.2, the $q$-lifted conformal block are given.
An expansion of the $q$-deformed conformal block in the cross ratio
is also mentioned.

The Kadell formula for the Macdonald polynomials is the basic calculational
tool of our study \cite{IO5}. The Kadell formula and their explicit forms for a few Macdonald
polynomials are summarized in subsection 4.3.


\subsection{Conformal block: From DF to Selberg}


Let us consider the conformal field theory with the central charge $c$
\be\label{ccharge}
c = (n-1) \bigl( 1 - n (n+1) Q_E^2 \bigr), \qq
Q_E = \sqrt{\beta} - \frac{1}{\sqrt{\beta}},
\ee
associated with the $A_{n-1} = \mathfrak{sl}(n)$ Lie algebra. 
Let $\mathfrak{h}$ be the Cartan subalgebra of $A_{n-1}$
and $\mathfrak{h}^*$ its dual.

The four-point conformal block of the chiral vertex operators 
can be expressed in terms of free fields.
Let $\phi(z)$ be an $\mathfrak{h}$-valued free chiral boson
with correlation functions:
\be
\begin{split}
\langle \phi_a(z)\phi_b(w) 
\rangle = (e_a,e_b) \log(z-w), ~~~
\phi_a(z) =  \langle e_a, \phi(z) \rangle,
\\
e_a \in \mathfrak{h}^* : ~ \text{a simple root of $A_{n-1}$},  ~~~
a= 1, \cdots, n-1.
\end{split}
\ee
Here $(\cdot, \cdot)$ is the symmetric bilinear form on
$\mathfrak{h}^*$
and $\langle \cdot, \cdot
\rangle$
is the natural pairing between $\mathfrak{h}^*$ and $\mathfrak{h}$.
With $(e_a, e_a)=2$,
$C_{ab} := (e_a, e_b)$ is the Cartan matrix of the $A_{n-1}$ algebra.
The corresponding energy-momentum tensor $T(z)$ is given by
\be
\begin{split}
T(z) &= \frac{1}{2} : \mathcal{K}\bigl( \partial \phi(z), 
\partial \phi(z) \bigr): + Q_E \langle \rho, \partial^2 \phi(z)
\rangle \cr
&= \sum_{a,b=1}^{n-1} ( C^{-1})^{ab}
\left(  \frac{1}{2} : \partial \phi_a(z) \partial \phi_b(z) :
+ Q_E \partial^2 \phi_b(z) \right).
\end{split}
\ee
Here $\mathcal{K}(\cdot, \cdot)$ is the Killing form and
$\rho$ is the half the sum of positive roots of the $A_{n-1}$ algebra:
\be
\rho = \frac{1}{2} \sum_{a=1}^{n-1} a ( n-a) e_a.
\ee
The scaling dimension of the vertex operator 
$V_{\alpha}(z)=:\ex^{\langle \alpha, \phi(z) \rangle}:$ $(\alpha \in
\mathfrak{h}^*)$ is given by
\be
\frac{1}{2} ( \alpha, \alpha) - Q_E ( \rho, \alpha).
\ee

The free-field representation of the conformal block is then given by
\be
\begin{split}
\label{ZDF}
& \mathcal{F}(c, \Delta_I, \Delta_i | \Lambda) \cr 
&= \biggl\langle
V_{(1/\sqrt{\beta})\alpha_1}(0)
V_{(1/\sqrt{\beta})\alpha_2}(\Lambda)
V_{(1/\sqrt{\beta})\alpha_3}(1)
V_{(1/\sqrt{\beta})\alpha_4}(\infty)
\prod_{a=1}^{n-1} \mathcal{Q}_a{}^{N_a} 
\widetilde{\mathcal{Q}}_a{}^{\widetilde{N}_a}
\biggl\rangle,
\end{split}
\ee
with $\alpha_i \in \mathfrak{h}^*$ $(i=1,2,3,4)$.
Two kinds of screening charges are inserted:
\be
\mathcal{Q}_a = \int_0^{\Lambda} \de z \, V_{\sqrt{\beta}
  e_a}(z), \qq
\widetilde{\mathcal{Q}}_a
= \int_1^{\infty} \de z\, V_{\sqrt{\beta} e_a}(z).
\ee
The four points are set to
$z_1=0$, $z_2=\Lambda$, $z_3=1$, $z_4=\infty$. 
Hence $\Lambda$ is the cross ratio.
The central charge $c$ is given by \eqref{ccharge},
the scaling dimensions $\Delta_i$ of the external states are given by
\be
\Delta_i = \frac{1}{2\beta} ( \alpha_i, \alpha_i)
- \left( 1 - \frac{1}{\beta} \right)  ( \rho, \alpha_i), \qq
(i=1,2,3,4),
\ee
and the scaling dimension $\Delta_I$ 
of the intermediate state is given by
\be
\Delta_I = \frac{1}{2\beta} ( \alpha_I, \alpha_I)
- \left( 1 -\frac{1}{\beta} \right) ( \rho, \alpha_I),
\ee
with
\be
\alpha_I := \alpha_1 + \alpha_2 + \beta \sum_{a=1}^{n-1} 
N_a e_a = - ( \alpha_3 + \alpha_4 ) - \beta
\sum_{a=1}^{n-1} \widetilde{N}_a e_a + 2(\beta-1)  \rho.
\ee
Here we have used the momentum conservation condition: 
\be
\frac{1}{\sqrt{\beta}}( \alpha_1 + \alpha_2 + \alpha_3 + \alpha_4)
+ \sqrt{\beta} \sum_{a=1}^{n-1} ( N_a + \widetilde{N}_a ) e_a
- 2 Q_E \, \rho = 0.
\ee 

By using the Wick's theorem,  
we obtain the Dotsenko-Fateev integral representation of the conformal block
\eqref{ZDF} \cite{DF}
\be
\begin{split}
\mathcal{F} &= \Lambda^{(\alpha_1, \alpha_2)/\beta}
(1 - \Lambda)^{(\alpha_2, \alpha_3)/\beta}  \cr
& \qq \times \int \de z 
\Bigl(  \Delta_{A_{n-1}}(z) \Bigr)^{\beta}
\prod_{I=1}^{N_a+ \widetilde{N}_a} (z_I^{(a)})^{(\alpha_1,e_a)}
|z_I^{(a)} - \Lambda|^{(\alpha_2,e_a)} |z_I^{(a)} -
1|^{(\alpha_3,e_a)},
\end{split}
\ee
where
\be
\int \de z = \prod_{a=1}^{n-1} 
\left\{ \prod_{I=1}^{N_a} \int_{0}^{\Lambda} \de z_I^{(a)} 
\prod_{J=N_a+ 1}^{N_a+\widetilde{N}_a} \int_1^{\infty} 
\de z_{J}^{(a)} \right\}, 
\ee
\be
\Delta_{A_{n-1}}(z) 
= \prod_{a=1}^{n-1} 
\prod_{I<J}^{N_a + \widetilde{N}_a} |z_J^{(a)} -
z_I^{(a)}|^{2} \, 
\prod_{b=1}^{n-2} 
\prod_{I=1}^{N_b + \widetilde{N}_b} 
\prod_{J=1}^{N_{b+1} +\widetilde{N}_{b+1}} |z_J^{(b+1)} -
z_I^{(b)}|^{-1}.
\ee

By changing the integration variables from 
$z_I^{(a)}$ ($I=1,2,\dotsc, N_a+\widetilde{N}_a$)
to $x_I^{(a)}$ ($I=1,2,\dotsc, N_a$) and $y_J^{(a)}$
$(J=1,2,\dotsc, \widetilde{N}_a)$,
defined by 
\begin{align}
 \Lambda x_I^{(a)} := z_I^{(a)}, ~~~~
 \frac{1}{y_J^{(a)}} := z_{N_a+J}^{(a)},
\end{align}
we obtain the following Selberg-type multiple integral \cite{IO5}:
\be
\mathcal{F} = \Lambda^{\Delta_I - \Delta_1 - \Delta_2}
(1 - \Lambda)^{(\alpha_2,\alpha_3)/\beta}
Z_{S}(\Lambda),
\ee
where 
\be\label{DFtoS}
\begin{split}
Z_{S}(\Lambda) &:= 
 \int_{[0,1]^{N}} \de x \, \Bigl( \Delta_{A_{n-1}}(x) \Bigr)^{\beta}
\prod_{a=1}^{n-1} \prod_{I=1}^{N_a} ( x_I^{(a)})^{(\alpha_1, e_a)}
( 1 - x_I^{(a)})^{(\alpha_2, e_a)} \cr
& \qq \times
\int_{[0,1]^{\widetilde{N}}} \de y \, \Bigl( \Delta_{A_{n-1}}(y) \Bigr)^{\beta}
\prod_{b=1}^{n-1} \prod_{J=1}^{N_b} ( y_J^{(b)})^{(\alpha_4, e_b)}
( 1 - y_J^{(b)})^{(\alpha_3, e_b)} \, F(x,y|\Lambda).
\end{split}
\ee
Here $N:=\sum_{a} N_a$, $\widetilde{N}:=\sum_{a} \widetilde{N}_a$,
\be\label{Fxy}
F(x,y|\Lambda)
:= \prod_{a=1}^{n-1} \left\{
\prod_{I=1}^{N_a} ( 1 - \Lambda x_I^{(a)})^{(\alpha_3, e_a)}
\prod_{J=1}^{\widetilde{N}_a} ( 1 - \Lambda
y_J^{(a)})^{(\alpha_2,e_a)} 
\right\} 
\prod_{a=1}^{n-1} \prod_{b=1}^{n-1} 
\prod_{I=1}^{N_a} \prod_{J=1}^{\widetilde{N}_b}
( 1 - \Lambda x_I^{(a)} y_J^{(b)} )^{\beta C_{ab}} ,
\ee 
\be
\de x = \prod_{a=1}^{n-1} \prod_{I=1}^{N_a}  \de x_I^{(a)}, \qq
\de y = \prod_{a=1}^{n-1} \prod_{J=1}^{\widetilde{N}_a} 
\de y_J^{(a)},
\ee
\be
\Delta_{A_{n-1}}(x)
= \prod_{a=1}^{n-1} \prod_{I<J}^{N_a} | x_I^{(a)} - x_J^{(a)} |^2
\prod_{b=1}^{n-2} \prod_{I=1}^{N_b} \prod_{J=1}^{N_{b+1}}
| x_J^{(b+1)} - x_I^{(b)}|^{-1},
\ee
\be
\Delta_{A_{n-1}}(y)
= \prod_{a=1}^{n-1} \prod_{I<J}^{\widetilde{N}_a}
| y_I^{(a)} - y_J^{(a)} |^2
\prod_{b=1}^{n-2} \prod_{I=1}^{\widetilde{N}_b}
\prod_{J=1}^{\widetilde{N}_{b+1}}
| y_J^{(b+1)} - y_I^{(b)}|^{-1}.
\ee

Strictly speaking, in order for the multiple integral \eqref{DFtoS}
to be well-defined,
the integration domains $[0,1]^N$ and $[0,1]^{\widetilde{N}}$
must be deformed properly.

Let us denote the $A_{n-1}$ Selberg integral \cite{war0708}\footnote{See also \cite{ZM}.} by
\be
\begin{split}
& I^{A_{n-1}}_{k_1,\dotsc, k_{n-1}}
(\kappa; \alpha_1, \dotsc, \alpha_{n-1}; \beta)  \cr
&= \int_{C_{\beta}^{k_1, \dotsc, k_{n-1}}[0,1]}
\de t \, \Bigl( \Delta_{A_{n-1}}(t) \Bigr)^{\beta}
\prod_{a=1}^{n-1} 
\prod_{I=1}^{k_a} ( t_I^{(a)})^{\kappa_a-1} ( 1 -
t_I^{(a)})^{\alpha_a-1},
\end{split}
\ee
where $\kappa_a-1 = \delta_{a,n-1}( \kappa-1)$, 
\be
\de t = \prod_{a=1}^{n-1} \prod_{I=1}^{k_a} \de t^{(a)}_I,
\qq
\Delta_{A_{n-1}}(t) = \prod_{a=1}^{n-1} \prod_{I<J}^{k_a}
| t_I^{(a)} - t_J^{(a)} |^2 
\prod_{b=1}^{n-2} \prod_{I=1}^{k_b} \prod_{J=1}^{k_{b+1}}
| t_J^{(b+1)} - t_I^{(b)}|^{-1}.
\ee
For the definition of properly deformed integration domain $C^{k_1,\dotsc,
  k_{n-1}}_{\beta}[0,1]$, see \cite{war0708}.
For 
\[0\leq k_1 \leq n_2 \leq \dotsm \leq k_{n-1}, k_0=k_n=0,\]
\[
\mathrm{Re}(\kappa)>0, \ \ \mathrm{Re}(\alpha_a) >0 \ \ (a=1,2,\dotsc,
n-1),
\ \ 
-\mathrm{min}\left\{ \frac{\mathrm{Re}(\kappa)}{k_{n-1}-1},
\frac{1}{k_{n-1}} \right\} < \mathrm{Re}(\beta) < \frac{1}{k_{n-1}},
\]
and
\[
- \frac{\mathrm{Re}(\alpha_a)}{k_a-k_{a-1}-1} <
\mathrm{Re}(\beta) <
\frac{\mathrm{Re}(\alpha_a+\dotsm+\alpha_b)}{b-a}, \qq
(1 \leq a \leq b \leq n-1),
\]
it holds that
\be
\begin{split}
& I^{A_{n-1}}_{k_1,\dotsm, k_{n-1}}(\kappa; \alpha_1, \dotsc,
  \alpha_{n-1}; \beta) \cr
&= \prod_{1\leq a\leq b\leq n-1}
\prod_{j=1}^{k_a-k_{a-1}} \frac{\Gamma(\alpha_a+\dotsm+\alpha_b
+ (j+a-b-1)\beta )}
{\Gamma(\kappa_b + \alpha_a+\dotsm + \alpha_b +
  (j+a-b+k_b-k_{b+1}-2)\beta)} \cr
& \times \prod_{a=1}^{n-1} \prod_{j=1}^{k_a}
\frac{\Gamma(\kappa_a + (j-k_{a+1} -1)\beta) \, \Gamma(j\beta)}
{\Gamma(\beta)}.
\end{split}
\ee

We also use the following notation for normalized
$A_{n-1}$ Selberg integral
\be
S^{A_{n-1}}_{k_1, \dotsc, k_{n-1}}(\kappa; \alpha_1, \dotsc,
\alpha_{n-1}; \beta)
:= \frac{1}{\mathrm{vol}(C^{k_1, \dotsc, k_{n-1}}_{\beta}[0,1])}
I^{A_{n-1}}_{k_1, \dotsc, k_{n-1}}(\kappa; \alpha_1, \dotsc, 
\alpha_{n-1}; \beta),
\ee
where
\be
\mathrm{vol}(C^{k_1, \dotsc, k_{n-1}}_{\beta}[0,1])=
\int_{C^{k_1, \dotsc, k_{n-1}}_{\beta}[0,1]} \de t.
\ee

Let $\omega_a$ be the fundamental weights: 
$(\omega_a, e_b^{\vee})= \delta_{ab}$. 
Here $e_b^{\vee} = 2 e_b/(e_b,e_b) = e_b$ are the 
simple coroots.
We choose two external
momenta $\alpha_1$ and $\alpha_4$ to be proportional to $\omega_{n-1}$:
\be\label{momc}
\alpha_1 = u_+ \, \omega_{n-1}, \qq
\alpha_4 = u_- \, \omega_{n-1}.
\ee

Note that in the integrand of \eqref{DFtoS},
only the polynomial $F(x,y|\Lambda)$ \eqref{Fxy} has
dependence on the cross ratio $\Lambda$. With the conditions
\eqref{momc},
$Z_S$ factorizes into product of two $A_{n-1}$ Selberg integrals
at $\Lambda=0$:
\be
\begin{split}
Z_S(0) &= 
S^{A_{n-1}}_{N_1,\dotsc, N_{n-1}}
(1+u_+; 1+v_{1+}, \dotsm, 1+v_{(n-1)+}; \beta) \cr
& \times S^{A_{n-1}}_{\widetilde{N}_1, \dotsc, \widetilde{N}_{n-1}}
(1+u_-; 1 + v_{1-}, \dotsm, 1+v_{(n-1)-}; \beta),
\end{split}
\ee
where
\be
v_{a+}:=( \alpha_2, e_a), \qq
v_{a-}:=( \alpha_3, e_a).
\ee 

Let us rewrite \eqref{DFtoS} as
\be
Z_{S}(\Lambda)
= Z_{s}(0) \,  \Bigl\langle\!\!\Bigl\langle
F(x, y|\Lambda)
\Bigr\rangle_{+} \Bigr\rangle_{-},
\label{CB:Selberg}
\ee
Here $\langle \dotsm \rangle_{+}$ (resp. $\langle \dotsm \rangle_{-}$)
is the average over
the $A_{n-1}$ Selberg integral
$I^{A_{n-1}}_{N_1,\dotsc, N_{n-1}}(1+u_+; \{ 1 + v_{a+} \}; \beta)$
(resp. $I^{A_{n-1}}_{\widetilde{N}_1, \dotsm, \widetilde{N}_{n-1}}
(1+u_-; \{ 1 + v_{a-}\}; \beta)$), normalized as
$\langle 1 \rangle_{\pm} = 1$.

The AGT relation implies that $Z_S(\Lambda)/Z_S(0)$ is equal to
the instanton part
of the Nekrasov partition function of $\mathcal{N}=2$ $SU(n)$ gauge theory with
$N_f = 2n$ fundamental matters.


\subsection{$\bm{q}$-deformed conformal block}


In order to study the five-dimensional version of AGT conjecture,
we need a $q$-deformed conformal block.

\subsubsection{$\bm{q}$-deformation}

There is a simple recipe \cite{War,MMSS} to obtain a $q$-deformation of the Selberg-type multiple integral.
For the Selberg integral, defined by 
\begin{align}
S_n(\beta_1,\beta_2, \beta) 
 = \left(\prod_{I=1}^n \int_0^1 \de x_I \right) 
   \prod_{I=1}^n x_I^{\beta_1-1} (1 - x_I)^{\beta_2-1} 
   \prod_{1 \leq I < J \leq n} (x_I - x_J)^{2\beta},
\end{align}
replace the integral by the definite $q$-integral
\be
 \int_0^1 \de x_I ~~ \to ~~ \int_0^1 \de_q x_I,  
\ee
and the following factors by their $q$-deformed counterparts
\begin{equation}
\begin{split} 
 (1 - x_I)^{\beta_2-1} ~~ &\to ~~ \prod_{k=0}^{\beta_2-1}(1 - q^k x_I), \cr
 \prod_{1 \leq I < J \leq n} (x_I - x_J)^{2\beta} ~~ &\to ~~ 
 \prod_{1 \leq I \neq J \leq n} \prod_{k=0}^{\beta-1} (x_I - q^k x_J). 
\end{split}
\end{equation}
Then we obtain the $q$-Selberg integral: 
\begin{align*}
 S_{n,q} \equiv S_n(\beta_1,\beta_2, \beta; q) = 
  \left(\prod_{I=1}^n \int_0^1 \de_q x_I \right)
  \prod_{I=1}^n x_I^{\beta_1-1} 
  \prod_{k=0}^{\beta_2-1}(1 - q^k x_I)
  \prod_{1 \leq I \neq J \leq n} 
  \prod_{k=1}^{\beta-1} (x_I - q^k x_J). 
\end{align*}
Here for simplicity we have assumed $\beta_2$ and $\beta$ are positive
integers. We can easily modify the above expression when these parameters
are not integers.

By a similar replacement, we obtain the $q$-deformation of the conformal block (\ref{CB:Selberg}):
\begin{align}
\label{ZSq}
Z_{S}^{(q)} = \left\langle\!\!\left\langle
\prod_{a=1}^{n-1} \left\{ \prod_{I=1}^{N_a} \prod_{i=0}^{v_{a-}-1}
 (1 - \Lambda x_I^{(a)} q^i) 
 \prod_{J=1}^{\widetilde{N}_a} \prod_{j=0}^{v_{a+}-1} 
 (1 - \Lambda y_J^{(a)} q^j)
 \right\} \times \right.\right. ~~~~~~~~~~~\cr \left.\left.
 \times \prod_{a,b=1}^{n-1} \prod_{\ell=0}^{\beta-1} 
 \prod_{I=1}^{N_a} \prod_{J=1}^{\widetilde{N}_a} 
 ( 1 - \Lambda x_I^{(a)} y_J^{(b)} q^{\ell})^{C_{ab}}
\right\rangle_{N+,q} ~ \right\rangle_{\widetilde{N}-,q},
\end{align}
where 
\begin{align}
\label{avNpmq}
\biggl\langle f(x) \biggl\rangle_{N\pm,q} 
 =  \frac{1}{S_{N,q}} \left( \prod_{I=1}^{N} \int_0^1 \de_qx_I \right)
 \prod_{I=1}^{N} x_I^{u_{a\pm}} \prod_{i=1}^{v_{a\pm}-1} (1 - x_I q^{i}) 
 \prod_{1 \leq I \neq J \leq N} 
 \prod_{i = 1}^{\beta-1} (x_I - q^{i} x_J) f(x)
\end{align}
Rearranging the integrand of eq.\eqref{ZSq}, we obtain
\begin{align}
\label{ZSq2}
 Z_S^{(q)} = \bigl\langle \bigl\langle I_S^{(q)} 
 \bigl\rangle_{N+,q} \bigl\rangle_{\widetilde{N}-,q},
\end{align}
where
\begin{align}
\label{IS}
I_{S}^{(q)} = 
\exp \left\{ - \sum_{k=1}^{\infty} [\beta]_{q^k} \frac{\Lambda^k}{k}
 \sum_{a=1}^{n} \left[
 \left( r_k^{(a)} + \frac{[v_{a+}]'_{q^k}}{[\beta]_{q^k}} \right)
 \left( \tilde{r}_k^{(a)} + \frac{[v_{a-}]'_{q^k}}{[\beta]_{q^k}} \right)
 - \frac{[v_{a+}]'_{q^k}}{[\beta]_{q^k}} \frac{[v_{a-}]'_{q^k}}{[\beta]_{q^k}}
 \right] \right\}. 
\end{align}
Here we have introduced the $q$-number
\be
 [a]_q := \frac{1 - q^a}{1 - q},
\ee
and
\begin{align*}
&[v_{a-}]'_{q^k} := - \sum_{s=1}^{a-1} [v_{s-}]_{q^k}, ~~~~
 [v_{(n-a)+}]'_{q^k} := \sum_{s=1}^{a} [v_{(n-s)+}]_{q^k},
\end{align*}
with
\[
 [v_{1-}]'_{q^k} = [v_{n+}]'_{q^k} = 0.
\]
Let $p_k^{(a)}$ (resp. $\tilde{p}_k^{(a)}$) be the $k$-th power sum of
$x_I^{(a)}$ (resp. $y_I^{(a)}$):
\be
p_k^{(a)} = \sum_{I=1}^{N_a}( x_I^{(a)} )^k, \qq
\tilde{p}_k^{(a)} = \sum_{I=1}^{\widetilde{N}_a} ( y_I^{(a)} )^k, \qq
(a=1,2,\dotsc, n-1).
\ee 
The polynomials $r^{(a)}_k=r^{(a)}_k(x)$
and $\tilde{r}^{(a)}_k= \tilde{r}^{(a)}_k(y)$ in \eqref{IS} are defined 
in terms of the power sum polynomials as
\be
r^{(a)}_k := p_k^{(a)} - p_k^{(a-1)}, \qquad
\tilde{r}^{(a)}_k := \tilde{p}_k^{(a)} - \tilde{p}_k^{(a-1)}, \qquad
(a=1,2, \dotsc, n),
\ee
with understanding
\be
p_k^{(0)} = p_k^{(n)} = \tilde{p}_k^{(0)} = \tilde{p}_k^{(n)} = 0.
\ee

\subsubsection{$\bm{\Lambda}$-expansion of the $\bm{q}$-deformed conformal block}

In order to compare with the five-dimensional Nekrasov partition function,
let us consider an expansion of the q-deformed conformal block \eqref{ZSq2}
in the cross ratio $\Lambda$.

Let $\lambda$ be a partition, $\lambda'$ be its conjugate and $a_{\lambda}(s)$
(resp. $\ell_{\lambda}(s)$) be the arm length (resp. leg length)
at $s=(i,j) \in \lambda$:
\be
 a_{\lambda}(s) = a_{\lambda}(i,j)= \lambda_i - j, \qq 
\ell_{\lambda}(s) = \ell_{\lambda}(i,j)= \lambda'_j - i.
\ee

Using the Cauchy identity for the Macdonald polynomials $P_{\lambda}(x; q, t)$
with two parameters $q$, $t$,
\begin{equation}
\begin{split} 
& \sum_{\l} \frac{C_{\l}}{C'_{\l}} P_{\l}(x;q,t) P_{\l}(y;q,t)   \cr
&= \prod_{I,J} \frac{(t x_I y_J;q)_{\infty}}{(x_Iy_J;q)_{\infty}} 
 = \prod_{I,J} \prod_{K=0}^{\beta-1} (1 - x_I y_J q^{K})^{-1} 
 = \exp \left\{\sum_{k=1}^{\infty} [\beta]_{q^k} 
  \frac{1}{k} p_k \tilde{p}_k \right\}, 
\end{split}
\end{equation}
where
\be
 \frac{C_{\l}}{C'_{\l}} 
 = \prod_{s\in\l} 
 \frac{[ a_{\lambda}(s) + \beta \ell_{\lambda}(s) + 1]_q}
 {[a_{\lambda}(s) + \beta \ell_{\lambda}(s) + \b]_q},
\ee
we obtain a $\Lambda$-expansion of the $q$-deformed conformal block into a
basis given by 
products of the Macdonald polynomials:
\begin{align}
\label{ZSqM}
Z_{S}^{(q)} = 
 \sum_{k=0}^{\infty} \Lambda^{k} 
 \sum_{\substack{\vec{Y}\\|\vec{Y}| = k}} \prod_{a=1}^{n} \frac{C_{Y_a}}{C'_{Y_a}}
 \left\langle \prod_{a=1}^{n} 
 P_{{Y_a}} \left( - r_{k}^{(a)} - \frac{[v_{a+}]'_{q^k}}{[\beta]_{q^k}} 
 \right) \right\rangle_{N+} \times \cr
 \times \left\langle \prod_{a=1}^{n}
 P_{{Y_a}} \left( \tilde{r}_k^{(a)} + \frac{[v_{a-}]'_{q^k}}{[\beta]_{q^k}} 
 \right) \right\rangle_{\widetilde{N}-}.
\end{align}
 

\subsection{Kadell formula}


Unfortunately, we have no formula (or conjecture) for the 
average of the general $n$ products of the Macdonald polynomials which appear in \eqref{ZSqM}\footnote{
For $\beta=1$, see \cite{ZM}.}. For the formula for the $2$-Macdonald average, see \cite{MMSS}.
But first few terms of the expansion can be explicitly evaluated by
using the Kadell formula. 

The Kadell formula \cite{kan96,kad97} 
for the Macdonald polynomials $P_{\lambda}(x; q, q^{\beta})$
is given by 
\begin{align}
\label{KadellMac}
 \frac{1}{S_q} \left( \prod_{I=1}^N \int_0^1 \de_q x_I \right) 
 P_{\l}(x; q, q^{\b})
 \prod_I x_I^u \prod_{k=0}^{v-1} (1 - q^k x_I) 
 \prod_{I \neq J} \prod_{k=0}^{\b-1} (x_I - q^k x_J) \cr
 = q^{W_{\l}(v,\b)} 
 \frac{[N\b, \l]_q [u + N \b + 1 - \b, \l]_q}{d_q(\l)[u + v + 2N\b + 2 - 2\b, \l]_q},
\end{align}
where 
\begin{align}
 &[x,\l]_{q} = \prod_{(i,j)\in\l} [x + \b(1-i) + (j-1)]_q, \\
 &d_q(\l) = \prod_{(i,j) \in \l} [\b + (\l_i -j) + \b (\l'_j -i)]_q 
              = \prod_{ s \in \l} [\beta +a_{\lambda} (s) + \beta \ell_{\lambda} (s)]_q, \\
 &q^{W_{\l}(v,\b)} = \prod_{(i,j)\in\l} q^{v + (i-1)\b}.
\end{align}

Using the average \eqref{avNpmq}, 
explicit forms of \eqref{KadellMac} for first few Macdonald polynomials 
are
(with rescaling $u_{a\pm} \to \b  u_{a\pm}, ~~ v_{a\pm} \to \b v_{a\pm}$)
\begin{align*}
 &\langle P_{(0)} \rangle_{N \pm, q} = 1, \\
 &\langle P_{(1)} \rangle_{N \pm, q} = 
   t^{v_{a\pm}} [N]_t 
   \frac{[u_{a\pm} + N  - (1-\b^{-1})]_{t}}{[w_{a\pm} - 2 (1-\b^{-1})]_{t}},\\
 &\langle P_{(2)} \rangle_{N \pm, q} =
   t^{2v_{a\pm}} [N]_t
   \left[ \frac{N\b + 1}{\b + 1} \right]_{t q} 
  \frac{[u_{a\pm} + N  - (1 - \b^{-1})]_{t}
             [u_{a\pm} + N  - (1- 2 \b^{-1} )]_{t}}
           {[w_{a\pm} -  (2 - 2 \b^{-1})]_{t} 
             [w_{a\pm} - (2 - 3\b^{-1})]_{t}},\\
 &\langle P_{(1^2)} \rangle_{N \pm, q}  = 
   t^{2v_{a\pm}+1} [N]_t  \left[\frac{N-1}{2} \right]_{t^2}
   \frac{[u_{a\pm} + N - (1 - \b^{-1})]_{t}
             [u_{a\pm} + N - (2 - \b^{-1}) ]_{t}}
           {[w_{a\pm} - (2  - 2 \b^{-1})]_{t} 
             [w_{a\pm} - (3 - 2 \b^{-1})]_{t}}, \\
 &\langle P_{(3)} \rangle_{N \pm, q}  = 
   t^{3v_{a\pm}}  
   [N]_t \left[ \frac{N\b + 1}{\b + 1} \right]_{t q} 
   \left[ \frac{N\b + 2}{\b + 2} \right]_{t q^2}   
   \prod _{k=1}^{3} \frac{[u_{a\pm} + N - (1 - k \b^{-1})]_{t}}
                                       {[w_{a\pm} - (2 - (k + 1) \b^{-1})]_{t}}, \\
 &\langle P_{(2,1)} \rangle_{N \pm, q}  = 
   t^{3v_{a\pm} + 1}
   [N]_t [N-1]_t \left[\frac{N\b + 1}{2\b+1}\right]_{t^2 q} 
  \prod_{k=1}^2 \frac{[u_{a\pm} + N - (1 - k\b^{-1})]_{t}}
                                  {[w_{a\pm} - (2 - (k+1)\b^{-1})]_{t}} \times \\
 &\hspace{9cm}  \times \frac{[u_{a\pm} + N - (2 - \b^{-1})]_{t}}
           {[w_{a\pm} - (3 - 2 \b^{-1})]_{t}},\\
 &\langle P_{(1^3)} \rangle_{N \pm, q}  = 
   t^{3v_{a\pm} + 3}
   [N]_t \left[ \frac{N-1}{2} \right]_{t^2} \left[ \frac{N-2}{3} \right]_{t^3}
   \prod _{k=1}^{3} \frac{[u_{a\pm} + N - (k - \b^{-1})]_{t}}
                                         {[w_{a\pm} - ( (k+1) - 2\b^{-1})]_{t}},
\end{align*}
where $t=q^{\beta}$ and $w_{a\pm} = u_{a\pm} + v_{a\pm} + 2 N$.

Remark. In the $q \to 1$ limit, \eqref{KadellMac}
turns into the Kadell formula for the Jack polynomials 
$P_{\lambda}^{(1/\beta)}(x)$:
\begin{align}
 \frac{1}{S} \left( \prod_{I=1}^N \int_0^1 \de x_I \right) 
 P_{\l}^{(1/\b)}(x)
 \prod_I x_I^u  (1 - x_I)^v 
 \prod_{I \neq J} (x_I -  x_J)^{2\b}\cr
 = 
 \frac{[N\b, \l] [u + N \b + 1 - \b, \l]}{d(\l)[u + v + 2N\b + 2 - 2\b, \l]},
\end{align}
where 
\begin{align}
 &[x,\l] = \prod_{(i,j)\in\l} [x + \b(1-i) + (j-1)] 
 = \prod_{i \geq 1} (x + \b(1-i))_{\l_i}, \\
 &d(\l) = \prod_{(i,j) \in \l} (\b + (\l_i -j) + (\l'_j -i)). 
\end{align}


\section{5d instanton partition function and its projection onto ALE space}



\subsection{brief review of 4d $\bm{SU(n)}$ instanton partition function on ALE}


The instanton partition function of four dimensional $\mathcal{N}=2$ 
supersymmetric $SU(n)$ gauge theory on $\mathbb{R}^4$ (with $\Omega$-deformation)
can be calculated by the method of localization. 
The fixed points of the torus action $U(1)^2\times U(1)^n$ 
 is labeled by an $n$-tuple of Young diagrams 
$\vec{Y}=(Y_{\a})_{\a=1,2,\cdots,n}$. 
For the fixed points corresponding to the $k$-instanton, 
$|\vec{Y}| = \sum_{\alpha} | Y_{\alpha}|= k$.  
Here we denote by $|Y_{\a}|$ the number of boxes carried by $Y_{\alpha}$. 
The instanton part of the Nekrasov function 
of the $SU(n)$ theory with $N_f=2n$ fundamental matters is given by 
\be
Z^{\mathbb{R}^4} 
 = \sum_{k=0}^{\infty} \Lambda^{k} 
  \sum_{|\vec{Y}|=k} {\mathcal{A}}_{\vec{Y}}, 
 \label{Nek:4d}
\ee
where 
${\mathcal{A}}_{\vec{Y}}$ represents the contribution of 
 the fixed point $\vec{Y}$, 
\be
 {\mathcal{A}}_{\vec{Y}} = \frac{\dis \prod_{\alpha=1}^n 
\prod_{k=1}^{2n} f_{Y_{\alpha}}(m_k + a_{\alpha})}
{\dis \prod_{\alpha_1,\alpha_2=1}^{n} g_{Y_{\alpha_1} Y_{\alpha_2}}
(a_{\alpha_2} - a_{\alpha_1})}.
\label{Nek:4dA}
\ee
Here 
\be
\begin{split}
 g_{Y_1 Y_2} (x) &= \prod_{(i,j) \in Y_1} \biggl(x + \beta
  \ell_{Y_1}(i,j) 
+ a_{Y_2}(i,j) 
 + \beta \biggl)
 \biggl( x + \beta \ell_{Y_1}(i,j) + a_{Y_2}(i,j) + 1 \biggl),  \cr
 f_{Y} (x) &=  \prod_{(i,j) \in Y} \biggl( x - \beta(i-1) + (j-1)
    \biggl),
 \end{split}
\ee
\be
 a_Y(i,j) = Y_i - j, ~~~~~
 \ell_Y(i,j) =  Y_j' - i. 
\ee
Here $\epsilon_1 = \sqrt{\beta} g_s$ and $\epsilon_2 = - g_s/\sqrt{\beta}$. 
Therefore, 
\be
 g_s^2 = - \epsilon_1 \epsilon_2, ~~~~~~
 \beta = - \frac{\epsilon_1}{\epsilon_2}. 
\ee
Suppose that the torus action is generated by 
$(\epsilon_1,\epsilon_2,a_1,\cdots,a_n)$. 
Then the weight of an individual box $(i,j) \in Y_{\alpha}$ of Young diagram is
given by $a_{\alpha} + (i-1) \epsilon_1 +(j-1) \epsilon_2$. 

Let us consider the instanton on $\mathbb{R}^4/\mathbb{Z}_r$. 
The $\mathbb{Z}_r$ action is \cite{KN,FMP}
\be
 \epsilon_1 \to \epsilon_1 - \frac{2 \pi \im}{r}, ~~~
 \epsilon_2 \to \epsilon_2 + \frac{2 \pi \im}{r}, ~~~
 a_{\alpha} \to a_{\alpha} + q_{\alpha} \frac{2\pi \im}{r}. 
 \label{shift:orbifold}
\ee From the weight associated with the torus action, 
we see that $\mathbb{Z}_r$-charges are assigned to each Young diagram and 
its boxes. 
The charge of each box $(i,j)$ of the Young diagram $Y_{\alpha}$ is 
\be
q_{\alpha,(i,j)} = q_{{\alpha}} - (i-1) + (j-1)  ~ ({\rm mod}~r), ~~~~ 
\alpha=1,2,\cdots,n, 
\ee
and $q_{{\alpha}}$ is regarded as the charge of each Young diagram $Y_{\alpha}$.


\subsection{more on the labeling of ALE instantons}


Let us consider the case where the 1st Chern class vanishes. 
This condition leads to
\be 
 n_{\ell} - 2 k_{\ell} + k_{{\ell}+1} + k_{{\ell}-1} = 0. 
 \label{cond:chern}
\ee 
Here $n_{\ell}$ is the number of Young diagram $\{Y_{\alpha}\}$ such that 
$\mathbb{Z}_r$-charge $q_{{\alpha}} = \ell$ and 
$k_{\ell}$ is the total number of the boxes such that the
 $\mathbb{Z}_r$-charge $q_{\alpha,(i,j)}=\ell$. 
Of course they satisfy $n=\sum_{\ell} n_{\ell}$ and $k=\sum_{\ell} k_{\ell}$. 
When (\ref{cond:chern}) is satisfied, the 2nd Chern number is $k/r$ and 
the ALE partition function is schematically written as  
\be
Z^{\mathbb{R}^4/\mathbb{Z}_r} 
 = \sum_{k=0}^{\infty} ({\Lambda'})^{k/r} 
  \sum_{|\vec{Y}|=k} \mathcal{A}_{\vec{Y}}
  \biggl|_{\substack{\text{1st Chern} = 0\\ 
  \mathbb{Z}_r\text{charge of each factor = 0}
  }}.  
\ee


\subsection{limiting procedure from 5d instanton partition function}


On the other hand, the five dimensional Nekrasov partition function is
 given by \cite{AK,AY}
\be
 Z^{\mathbb{R}^5} 
 = \sum_{k=0}^{\infty} \widetilde{\Lambda}^k 
  \sum_{|\vec{Y}|=k} \widetilde{\mathcal{A}}_{\vec{Y}}, 
  \label{Nekrasov:5dim}
\ee
\be
 \widetilde{\mathcal{A}}_{\vec{Y}} =  \frac{\prod_{\alpha=1}^n \prod_{k=1}^{n} 
            f^{q+}_{Y_{\alpha}}(m_k + a_{\alpha}) 
            f^{q-}_{Y_{\alpha}}(m_{k+n} + a_{\alpha})}
            {\prod_{\alpha,\alpha'}^{n} g^q_{Y_{\alpha}Y_{\alpha'}}
            (a_{\alpha} - a_{\alpha'})}, 
\ee
\begin{align}
 & g^q_{YW}(x) = \prod_{(i,j) \in Y} 
    [x + \beta \ell_{Y}(i,j) + a_W(i,j) + \beta]_q
    [- x - \beta \ell_{Y}(i,j) - a_W(i,j) - 1]_q,   
    \label{Nek:gauge}\\
 & f^{q\pm}_{Y}(x) 
    = \prod_{(i,j) \in Y} 
   [ \pm x \mp \beta(i-1)  \pm (j-1)]_q, 
    ~~~~~~~~~~~~~~~ 
 [x]_q = \frac{1-q^x}{1-q}. 
 \label{Nek:fund}
\end{align}
The parameter $m_i~(i=1, \cdots, 2n)$ 
 is related to the five dimensional fundamental mass $m_i^{5d}$ by  
\be
m_i = m_i^{5d} + \frac{1}{2} (1 - \beta).
\ee
Eq. (\ref{shift:orbifold}) can be read off as the shift in $q$ 
 as well as in $t$ \cite{kim}
\be
 q = \ex^{\epsilon_2} \to \omega \, q, ~~~~ t = \ex^{- \epsilon_1 } \to \omega \, t,~~~~ 
 q^{\frac{a_{\alpha}}{\epsilon_2}} 
 = \ex^{a_{\alpha}} \to \omega^{q_{\alpha}} q^{\frac{a_{\alpha}}{\epsilon_2}}, 
 ~~~~~~~~~~~~ \omega=\ex^{2 \pi \im/r}. 
 \label{shift:q,t}
\ee
Here we have rescaled $a_{\alpha} \to a_{\alpha}/\epsilon_2$. 
If we take the limit $q \to 1$ subsequently, we expect that this is equivalent to 
taking the $\mathbb{Z}_r$ orbifold projection on four dimensional space. 
On the other hand, this limit is equal to the root of unity limit of $q$ and $t$ and 
the five dimensional Nekrasov partition function reduces to 
 that on $\mathbb{R}^4/\mathbb{Z}_r$. 
In what follows, let us realize this procedure as
\be
 q = \omega \, \ex^{h \epsilon_2}, ~~~ t = \omega \, \ex^{-h \epsilon_1}, ~~~ 
 q^{\frac{a_{\alpha}}{\epsilon_2}} = \omega^{q_{\alpha}} \, \ex^{ h a_{\alpha}}, ~~~
 q^{\frac{m_{i}}{\epsilon_2}} = \ex^{h m_{i}}, ~~~~~
 h \to +0. 
 \label{limit:gauge}
\ee
Here we have rescaled $m_i$ by $m_i/\epsilon_2$.

Taking the limit (\ref{limit:gauge}), 
the leading term in the expansion of (\ref{Nek:gauge}) around $h=0$ is given by  
\begin{align}
 &g_{Y_{\alpha}Y_{\alpha'}}^q (a_{\alpha'}/\epsilon_2 - a_{\alpha}/\epsilon_2)  \cr
 &= \prod_{(i,j) \in Y_{\alpha}} \left[\frac{a_{{\alpha'\alpha}}}{\epsilon_2} 
 + \beta \ell_{Y_{\alpha}}(i,j) + a_{Y_{\alpha'}}(i,j) + \beta \right]_q 
 \left[-\frac{a_{\alpha'\alpha}}{\epsilon_2} 
 - \beta \ell_{Y_{\alpha}}(i,j) - a_{Y_{\alpha'}}(i,j) - 1 \right]_q \cr
 &= \prod_{\substack{(i,j) \in Y_{\alpha} \\ 
 q_{{\alpha'\alpha}} + \ell_{Y_{\alpha}}(i,j) + a_{Y_{\alpha'}}(i,j) + 1 \neq 0 ({\rm mod}~r)}}
 \frac{1 - \omega^{q_{\alpha'\alpha} + \ell_{Y_{\alpha}}(i,j) + a_{Y_{\alpha'}}(i,j) + 1 }}{1 - \omega} 
 \frac{1 - \omega^{-q_{\alpha'\alpha} - \ell_{Y_{\alpha}}(i,j) - a_{Y_{\alpha'}}(i,j) - 1 }}{1 - \omega} \cr
 & \prod_{\substack{(i,j) \in Y_{\alpha} \\ 
 q_{\alpha'\alpha} + \ell_{Y_{\alpha}}(i,j) + a_{Y_{\alpha'}}(i,j) + 1 = 0 ({\rm mod}~r)}}
 \left( \frac{- h}{1 - \omega}\right)^2  
 \biggl( a_{\alpha'\alpha} 
 - \epsilon_1 \ell_{Y_{\alpha}}(i,j) +\epsilon_2 a_{Y_{\alpha'}}(i,j) 
 - \epsilon_1 \biggl) \cr
 & \hspace{6cm}
 \times \biggl(-a_{\alpha'\alpha}
 + \epsilon_1 \ell_{Y_{\alpha}}(i,j) - \epsilon_2 a_{Y_{\alpha'}}(i,j) 
 - \epsilon_2 \biggl). 
\label{Nek:g}
\end{align}
Here we have set $a_{\alpha'\alpha} = a_{\alpha'} - a_{\alpha}$ and
 $q_{\alpha'\alpha} = q_{\alpha'} - q_{\alpha}$. 
Similarly, from (\ref{Nek:fund}) we obtain 
\begin{align}
 &f_{Y_{\alpha}}^{q\pm} (m_i/\epsilon_2 + a_{s}/\epsilon_2)  \cr
 &= \prod_{(i,j) \in Y_{\alpha}} \left[
 \pm \left( \frac{m_i}{\epsilon_2} + \frac{a_{\alpha}}{\epsilon_2} \right)
 \mp  \beta(i-1)  \pm (j-1)  \right]_q \cr
 &= \prod_{\substack{(i,j) \in Y_{\alpha} \\ 
 \pm q_{\alpha} \mp (i-1) \pm (j-1) \neq 0 ({\rm mod}~r)}}
 \frac{1-\omega^{\pm q_{\alpha} \mp (i-1) \pm (j-1) }}{1-\omega}
 \cr
 & \prod_{\substack{(i,j) \in Y_{\alpha} \\ 
 \pm q_{\alpha} \mp (i-1) \pm (j-1) = 0 ({\rm mod}~r)}}
 \frac{- h}{1 - \omega}
 \biggl(
 \pm \left( m_i + a_{\alpha} \right)
 \pm  \epsilon_1 (i-1)  \pm \epsilon_2 (j-1)
 \biggl).
\label{Nek:f}
\end{align}
The above results show that 
among the factors which 
 compose the five dimensional Nekrasov partition function, 
 only the factors which 
 contribute to the $\mathbb{R}^4/\mathbb{Z}_r$ instanton partition function 
 survive the limiting procedure (\ref{limit:gauge}) at the leading order in $h$. 
The other factors become automatically unimportant coefficients. 

\begin{figure}[h]
\begin{center}
  \includegraphics[height=2cm]{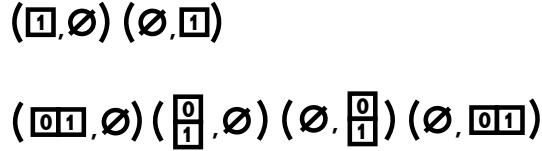}
\end{center}
\caption{Samples of Young diagrams contributing in the case of $SU(2)$ and $r=2$. 
The upper corresponds to $k_0=0$ of case 1 and the lower corresponds to 
$k_0=1$ of case 2. 
The number drawn in each box is the respective $\mathbb{Z}_r$-charge.
\label{Young:ALE}
}
\end{figure}

In what follows, let us first consider the case of $SU(2)$ and $r=2$. 
Setting $a \equiv a_1=-a_2$, we must have the charge $q_a=q_{1}=-q_{2}$.  From 
 the condition (\ref{cond:chern}), the following two cases are permitted: 
\begin{align} 
 &\text{case 1}:~ q_a=1, ~~~ (k_0,k_1)=(k_0,k_0+1), 
 \label{case1:r=2} \\
 &\text{case 2}:~ q_a=0, ~~~ (k_0,k_1)=(k_0,k_0). 
 \label{case2:r=2}
\end{align}  
Here $k_0=0,1,2,\cdots$. 
Therefore, we have (Fig. \ref{Young:ALE})
\begin{align}
 &{\rm case~1:}~ q = -\ex^{h \epsilon_2}, ~~~ t = - \ex^{-h \epsilon_1}, ~~~ 
 q^{a} = - \ex^{h a}, \\
 &{\rm case~2:}~ q = -\ex^{h \epsilon_2}, ~~~ t = - \ex^{-h \epsilon_1}, ~~~ 
 q^{a} = \ex^{h a}. 
\end{align}

Let us consider 
\begin{align}
 &Z^{(2)}_{k_0+1/2} 
 := \lim_{h \to 0} \frac{h^2}{2^2} \sum_{|A|+|B|=2k_0+1} \widetilde{\mathcal{A}}_{AB}^{(1,1)}, 
 \label{Nek1:r=2_limit}\\ 
 &Z^{(2)}_{k_0} := \lim_{h \to 0} \sum_{|A|+|B|=2k_0} \widetilde{\mathcal{A}}_{AB}^{(0,0)}. 
 \label{Nek0:r=2_limit} 
\end{align} 
Here $\mathcal{A}_{AB}^{(x,y)}$ represents that two  
 Young diagrams $A$ and $B$ have the $\mathbb{Z}_2$-charge $x$ and $y$ respectively. 
The coefficient $h^2/2^2$ in the definition of $Z^{(2)}_{k_0+1/2}$ 
 has introduced in order to remove 
the  coefficients in (\ref{Nek:g}) and (\ref{Nek:f}).   From (\ref{Nek0:r=2_limit}) and (\ref{Nek1:r=2_limit}), 
we will now check that the $\mathbb{R}^4/\mathbb{Z}_2$ partition function is 
\be
 Z^{\mathbb{R}^4/\mathbb{Z}_2}_{SU(2)} 
 = \sum_{k_0=0}^{\infty} \left( {\Lambda'} \right)^{k_0} Z^{(2)}_{k_0} 
 + \sum_{k_0=0}^{\infty} 
 \left( {\Lambda}' \right)^{k_0+1/2} Z^{(2)}_{k_0+1/2},  
\ee
by the explicit calculation. 
First of all, in the case $k_0=0~(k=1)$ of case 1, 
expanding around $h=0$, we have 
\be
 \widetilde{\mathcal{A}}_{(1)(0)}^{(1,1)} = \frac{2^2}{h^2} 
 \left( \frac{- 1}{2a(2a - \epsilon_1 - \epsilon_2) } + O(h) \right), ~~
 \widetilde{\mathcal{A}}_{(0)(1)}^{(1,1)} = \frac{2^2}{h^2} 
 \left( \frac{- 1}{2a(2a + \epsilon_1 + \epsilon_2) }  + O(h) \right). 
\ee 
Therefore, 
\be
 Z^{(2)}_{1/2} 
 = \lim_{h \to 0}\frac{h^2}{2^2} 
  (\widetilde{\mathcal{A}}_{(1)(0)}^{(1,1)} + \widetilde{\mathcal{A}}_{(0)(1)}^{(1,1)}) 
 = - \frac{2}{(2a+\epsilon_1+\epsilon_2)(2a-\epsilon_1-\epsilon_2)}. 
\ee 
Next, in the case $k_0=1~(k=2)$ of case 2, 
\be
\begin{split}
 \widetilde{\mathcal{A}}_{(2),(0)} &= 
 \frac{\prod_{i=1}^2[m_i+a]_q[-(m_{i+2}+a)]_q[m_i+a+1]_q[-(m_{i+2}+a+1)]_q}
 {[1+\beta]_q[-2]_q[\beta]_q[-1]_q[-2a-1+\beta]_q[2a]_q[-2a-2+\beta]_q[2a+1]_q}, \\
 \widetilde{\mathcal{A}}_{(1,1),(0)} &= 
 \frac{\prod_{i=1}^2[m_i+a]_q[-(m_{i+2}+a)]_q[m_i+a-\beta]_q[-(m_{i+2}+a-\beta)]_q}
 {[2\beta]_q[-1-\beta]_q[\beta]_q[-1]_q[-2a+2\beta-1]_q
 [2a-\beta]_q[-2a-1+\beta]_q[2a]_q}, \\
 \widetilde{\mathcal{A}}_{(1),(1)} &= 
 \frac{\prod_{i=1}^2[m_i+a]_q[-(m_{i+2}+a)]_q[m_i-a]_q[-(m_{i+2}-a)]_q}
 {[\beta]_q[-1]_q[\beta]_q[-1]_q[-2a+\beta]_q[2a-1]_q[2a+\beta]_q[-2a-1]_q}, \\
 \widetilde{\mathcal{A}}_{(0),(1,1)} &= 
 \frac{\prod_{i=1}^2[m_i-a]_q[-(m_{i+2}-a)]_q[m_i-a-\beta]_q[-(m_{i+2}-a-\beta)]_q}
 {[2\beta]_q[-1-\beta]_q[\beta]_q[-1]_q[2a+2\beta-1]_q[-2a-\beta]_q
 [2a-1+\beta]_q[-2a]_q},\\
 \widetilde{\mathcal{A}}_{(0),(2)} &= 
 \frac{\prod_{i=1}^2[m_i-a]_q[-(m_{i+2}-a)]_q[m_i-a+1]_q[-(m_{i+2}-a+1)]_q}
 {[1+\beta]_q[-2]_q[\beta]_q[-1]_q[2a-1+\beta]_q[-2a]_q[2a-2+\beta]_q[-2a+1]_q}.
\end{split}
\ee
Assigning the charge (0,0), we obtain
\be  
\begin{split}
 \widetilde{\mathcal{A}}_{(2)(0)}^{(0,0)} &= 
 \frac{\prod_{i=1}^4(m_i+a)}
 {(-2\epsilon_2)(\epsilon_1-\epsilon_2)(2a+\epsilon_1+\epsilon_2)2a} + O(h), \\
 \widetilde{\mathcal{A}}_{(1,1)(0)}^{(0,0)} &= 
 \frac{\prod_{i=1}^4(m_i+a)}
 {2\epsilon_1(\epsilon_1-\epsilon_2)(2a+\epsilon_1+\epsilon_2)2a} + O(h), \\
 \widetilde{\mathcal{A}}_{(1)(1)}^{(0,0)} &= 
 \frac{h^8}{2^8} \prod_{i=1}^4 (m_i + a)(m_i - a) + O(h^9), \\
 \widetilde{\mathcal{A}}_{(0)(1,1)}^{(0,0)} &=  
 \frac{\prod_{i=1}^4(m_i-a)}
 {2\epsilon_1(\epsilon_1-\epsilon_2)(2a-\epsilon_1-\epsilon_2))2a} + O(h),  \\
 \widetilde{\mathcal{A}}_{(0)(2)}^{(0,0)} &= 
 \frac{\prod_{i=1}^4(m_i-a)}
 {(-2\epsilon_2)(\epsilon_1-\epsilon_2)(2a-\epsilon_1-\epsilon_2)2a} + O(h). 
\end{split}
\ee
and 
\begin{align}
 Z^{(2)}_{1} 
 &=\lim_{h\to 0} 
 \left(\widetilde{\mathcal{A}}_{(2)(0)}^{(0,0)}
 + \widetilde{\mathcal{A}}_{(1,1)(0)}^{(0,0)} 
 + \widetilde{\mathcal{A}}_{(1)(1)}^{(0,0)} 
 + \widetilde{\mathcal{A}}_{(0)(1,1)}^{(0,0)} 
 + \widetilde{\mathcal{A}}_{(0)(2)}^{(0,0)} \right) \cr
 &=
 -\frac{\prod_{i=1}^4(m_i+a)}{\epsilon_1\epsilon_2(2a+\epsilon_1+\epsilon_2)4a} 
 - \frac{\prod_{i=1}^4(m_i-a)}{\epsilon_1\epsilon_2(2a-\epsilon_1-\epsilon_2)4a}.
\end{align}
Note that $\widetilde{\mathcal{A}}_{(1)(1)}^{(0,0)}$ which 
 does not satisfy (\ref{cond:chern}) becomes automatically 0 in this limiting procedure. 

Similarly, in the case $k=3$ and $k=4$ we obtain the following results, 
\begin{align} 
 Z^{(2)}_{3/2} 
 &\scriptstyle{ 
 =  \underset{h \to 0}{\lim} 
 \frac{h^2}{2^2} (\widetilde{\mathcal{A}}_{(3)(0)}^{(1,1)} + \widetilde{\mathcal{A}}_{(2,1)(0)}^{(1,1)} 
  + \widetilde{\mathcal{A}}_{(1,1,1)(0)}^{(1,1)}
  + \widetilde{\mathcal{A}}_{(2)(1)}^{(1,1)} + \widetilde{\mathcal{A}}_{(1,1)(1)}^{(1,1)} + 
 \widetilde{\mathcal{A}}_{(0)(3)}^{(1,1)} + \widetilde{\mathcal{A}}_{(0)(2,1)}^{(1,1)} + \widetilde{\mathcal{A}}_{(0)(1,1,1)}^{(1,1)} 
 + \widetilde{\mathcal{A}}_{(1)(2)}^{(1,1)} + \widetilde{\mathcal{A}}_{(1)(1,1)}^{(1,1)} ) }                \cr
 &\scriptstyle{
 = \frac{\prod_{i=1}^4 (m_i + a + \epsilon_2)}
 {(-\epsilon_1 + \epsilon_2)(-2\epsilon_2)(-2a - \epsilon_1 - \epsilon_2)(2a)(-2a-\epsilon_1-3\epsilon_2)(2a+2\epsilon_2)}
 + \frac{\prod_{i=1}^4 (m_i + a + \epsilon_1)}
 {(-2\epsilon_1)(\epsilon_1-\epsilon_2)(-2a-3\epsilon_1-\epsilon_2)(2a+2\epsilon_1)(-2a-\epsilon_1-\epsilon_2)(2a)}
 }                          \cr
 &~~\scriptstyle{
 + \frac{\prod_{i=1}^4 (m_i + a + \epsilon_2)}
 {(-\epsilon_1+\epsilon_2)(-2\epsilon_2)(-2a-\epsilon_1-\epsilon_2)(2a)(2a-\epsilon_1+\epsilon_2)(-2a-2\epsilon_2)}
 }
 \scriptstyle{
 + \frac{\prod_{i=1}^4 (m_i + a + \epsilon_1)}
 {(-2\epsilon_1)(\epsilon_1-\epsilon_2)(-2a-2\epsilon_1)(2a+\epsilon_1-\epsilon_2)(-2a-\epsilon_1-\epsilon_2)(2a)}
 }                      \cr
 &~~\scriptstyle{
 + \frac{\prod_{i=1}^4 (m_i - a + \epsilon_2)}
 {(-\epsilon_1 + \epsilon_2)(-2\epsilon_2)(2a - \epsilon_1 - \epsilon_2)(2a)(2a-3\epsilon_2-\epsilon_1)(-2a+2\epsilon_2)}
 + \frac{\prod_{i=1}^4 (m_i - a + \epsilon_1)}
 {(-2\epsilon_1)(\epsilon_1-\epsilon_2)(2a-3\epsilon_1-\epsilon_2)(-2a+2\epsilon_1)(2a-\epsilon_1-\epsilon_2)(-2a)}
 }                    \cr
 &~~\scriptstyle{
 + \frac{\prod_{i=1}^4 (m_i - a + \epsilon_2)}
 {(-\epsilon_1+\epsilon_2)(-2\epsilon_2)(2a-\epsilon_1-\epsilon_2)(-2a)(-2a-\epsilon_1+\epsilon_2)(2a-2\epsilon_2)}
 }
 \scriptstyle{
 + \frac{\prod_{i=1}^4 (m_i - a + \epsilon_1)}
 {(-2\epsilon_1)(\epsilon_1-\epsilon_2)(2a-2\epsilon_1)(-2a+\epsilon_1-\epsilon_2)(2a-\epsilon_1-\epsilon_2)(-2a)}
 }, 
\end{align}
and 
\begin{align} 
Z^{(2)}_{2} 
 &= \scriptstyle{
 \frac{\prod_{i=1}^4 (m_i + a)(m_i+a+2\epsilon_2)}
 {(-\epsilon_1+3\epsilon_2)(-4\epsilon_2)(-\epsilon_1+\epsilon_2)(-2\epsilon_2)(-2a-\epsilon_1-\epsilon_2)(2a)
 (-2a-\epsilon_1-3\epsilon_2)(2a+2\epsilon_2)}
 }                    \cr 
 &~~~\scriptstyle{
 + \frac{\prod_{i=1}^4 (m_i + a)(m_i+a+2\epsilon_2)}
 {(-\epsilon_1+\epsilon_2)(\epsilon_1-3\epsilon_2)(-\epsilon_1+\epsilon_2)(-2\epsilon_2)(-2a-\epsilon_1-3\epsilon_2)
 (2a+2\epsilon_2)(-2a-\epsilon_1-\epsilon_2)(2a)}
 }                    \cr
 &~~~\scriptstyle{
 + \frac{\prod_{i=1}^4 (m_i + a)(m_i+a+\epsilon_1+\epsilon_2)}
 {(-2\epsilon_1)(\epsilon_1-\epsilon_2)(-\epsilon_1+\epsilon_2)(-2\epsilon_2)(-2a-2\epsilon_1-2\epsilon_2)(2a+\epsilon_1+\epsilon_2)
 (-2a-\epsilon_1-\epsilon_2)(2a)}
 }                   \cr
 &~~~\scriptstyle{ 
 + \frac{\prod_{i=1}^4 (m_i + a)(m_i+a+2\epsilon_1)}
 {(-3\epsilon_1+\epsilon_2)(2\epsilon_1-2\epsilon_2)(-2\epsilon_1)(\epsilon_1-\epsilon_2)(-2a-3\epsilon_1-\epsilon_2)(2a+2\epsilon_1)
 (-2a-\epsilon_1-\epsilon_2)(2a)}
 }\cr
 &~~~\scriptstyle{ 
 + \frac{\prod_{i=1}^4 (m_i + a)(m_i+a+2\epsilon_1)}
 {(-4 \epsilon_1)(3\epsilon_1-\epsilon_2)(-2\epsilon_1)(\epsilon_1-\epsilon_2)(-2a-3\epsilon_1-\epsilon_2)(2a+2\epsilon_1)
 (-2a-\epsilon_1-\epsilon_2)(2a)}
 }                   \cr
 &~~~\scriptstyle{ 
 + \frac{\prod_{i=1}^4 (m_i + a)(m_i-a)}
 {(-2a-2\epsilon_1)(2a+\epsilon_1-\epsilon_2)(-2a-\epsilon_1-\epsilon_2)(2a)(-2a-\epsilon_1-\epsilon_2)(2a)
 (2a-\epsilon_1+\epsilon_2)(-2a-2\epsilon_2)}
 }                  \cr
 &~~~\scriptstyle{ 
 + \frac{\prod_{i=1}^4 (m_i + a)(m_i-a)}
 {(-\epsilon_1+\epsilon_2)(-2\epsilon_2)(-2\epsilon_1)(\epsilon_1-\epsilon_2)(-2a-\epsilon_1-\epsilon_2)(2a)(2a-\epsilon_1-\epsilon_2)(-2a)}
 }                  \cr
 &~~~\scriptstyle{ 
 + \frac{\prod_{i=1}^4 (m_i + a)(m_i-a)}
 {(\epsilon_2-\epsilon_1)^2(-2\epsilon_2)^2(-2a-\epsilon_1+\epsilon_2)(2a-2\epsilon_2)(2a-\epsilon_1+\epsilon_2)(-2a-2\epsilon_2)}
 }\cr
 &~~~\scriptstyle{ 
 + \frac{\prod_{i=1}^4 (m_i + a)(m_i-a)}
 {(\epsilon_1-\epsilon_2)^2(-2\epsilon_1)^2(-2a+\epsilon_1-\epsilon_2)(2a-2\epsilon_1)(2a+\epsilon_1-\epsilon_2)(-2a-2\epsilon_1)}
 }\cr
 &~~~\scriptstyle{ 
 + \frac{\prod_{i=1}^4 (m_i + a)(m_i+a+2\epsilon_2)}
 {(-\epsilon_1+3\epsilon_2)(-4\epsilon_2)(-\epsilon_1+\epsilon_2)(-2\epsilon_2)(2a-\epsilon_1-\epsilon_2)(-2a)
 (2a-\epsilon_1-3\epsilon_2)(-2a+2\epsilon_2)}
 }                    \cr 
 &~~~\scriptstyle{
 + \frac{\prod_{i=1}^4 (m_i + a)(m_i+a+2\epsilon_2)}
 {(-\epsilon_1+\epsilon_2)(\epsilon_1-3\epsilon_2)(-\epsilon_1+\epsilon_2)(-2\epsilon_2)(2a-\epsilon_1-3\epsilon_2)
 (-2a+2\epsilon_2)(2a-\epsilon_1-\epsilon_2)(-2a)}
 }                    \cr
 &~~~\scriptstyle{
 + \frac{\prod_{i=1}^4 (m_i + a)(m_i+a+\epsilon_1+\epsilon_2)}
 {(-2\epsilon_1)(\epsilon_1-\epsilon_2)(-\epsilon_1+\epsilon_2)(-2\epsilon_2)(2a-2\epsilon_1-2\epsilon_2)(-2a+\epsilon_1+\epsilon_2)
 (2a-\epsilon_1-\epsilon_2)(-2a)}
 }                   \cr
 &~~~\scriptstyle{ 
 + \frac{\prod_{i=1}^4 (m_i + a)(m_i+a+2\epsilon_1)}
 {(-3\epsilon_1+\epsilon_2)(2\epsilon_1-2\epsilon_2)(-2\epsilon_1)(\epsilon_1-\epsilon_2)(2a-3\epsilon_1-\epsilon_2)(-2a+2\epsilon_1)
 (2a-\epsilon_1-\epsilon_2)(-2a)}
 }\cr
 &~~~\scriptstyle{ 
 + \frac{\prod_{i=1}^4 (m_i + a)(m_i+a+2\epsilon_1)}
 {(-4 \epsilon_1)(3\epsilon_1-\epsilon_2)(-2\epsilon_1)(\epsilon_1-\epsilon_2)(2a-3\epsilon_1-\epsilon_2)(-2a+2\epsilon_1)
 (2a-\epsilon_1-\epsilon_2)(-2a)}
 }                   \cr
 &~~~\scriptstyle{ 
 + \frac{\prod_{i=1}^4 (m_i + a)(m_i-a)}
 {(2a-2\epsilon_1)(-2a+\epsilon_1-\epsilon_2)(2a-\epsilon_1-\epsilon_2)(-2a)(2a-\epsilon_1-\epsilon_2)(-2a)
 (-2a-\epsilon_1+\epsilon_2)(2a-2\epsilon_2)}
 }                  \cr
 &~~~\scriptstyle{ 
 + \frac{\prod_{i=1}^4 (m_i + a)(m_i-a)}
 {(-\epsilon_1+\epsilon_2)(-2\epsilon_2)(-2\epsilon_1)(\epsilon_1-\epsilon_2)(2a-\epsilon_1-\epsilon_2)(-2a)(-2a-\epsilon_1-\epsilon_2)(2a)}
 }.
\end{align}
If we consider the limit in which the fundamental matters decouple, 
the pure Yang-Mills case is reproduced. \cite{BMT}

\begin{figure}[h]
\begin{center}
  \includegraphics[height=2cm]{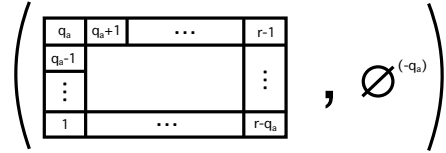}
\end{center}
\caption{Example of $n$-tuple Young diagram in the case of $k_0=0$, 
$SU(2)$ and general $r$. 
\label{young1}
}
\end{figure}

In the case of general $r$, setting $k_{\ell} = k_0 + \delta k_{\ell}$, 
~${\ell} = 0, \cdots, r-1 $, we find from the condition (\ref{cond:chern}), 
\begin{align}
 &q_a = 0 ~~~
 \delta k_{\ell} = (0, \cdots, 0), \\
 &q_a \neq 0 ~~~
 \delta k_{\ell} = (0, 1, 2, \cdots, \underset{{\ell}=q_a}{q_a},  \cdots, 
 \underset{{\ell}=r-q_a}{q_a}, q_a-1, \cdots, 1). 
\end{align}
Here 
\be
 0 \leq q_a \leq r- q_a ~~~~~ \Rightarrow ~~~~~ 
 0 \leq q_a \leq \left\lfloor \frac{r}{2} \right\rfloor. 
\ee
The total number of boxes in the case of $\mathbb{Z}_r$ charge $q_a$ 
is $rk_0 + q_a(r-q_a)$. 
Therefore, we define 
\be 
 Z^{(r)}_{k_0+ \frac{q_a(r-q_a)}{r}} 
 := \lim_{h \to 0} \Xi_{q_a} 
 \sum_{\substack{|A|+|B|\\=rk_0+q_a(r-q_a)}} \widetilde{\mathcal{A}}_{AB}^{(q_a,-q_a)}.  
\ee 
Now if we assume that 
$\Xi_{q_a}$ does not depend on $k$ and is equal to each other 
 for all $\widetilde{\mathcal{A}}_{AB}$ involved we are able to evaluate this quantity 
 for a typical Young diagram. 
At least, we can see that this assumption is correct at the lower levels of $SU(2)$, $r=2$. 
One of the case $k_0=0$ is drawn in Fig. \ref{young1}.
After the computation on this diagram, we find that we should set
\begin{align}
 &\Xi_0 = 1, ~~~~
 \Xi_1 = h^{2} \frac{1}{( 1 - \omega)(1-\omega^{-1})} , 
 \\
 &\Xi_{q_a} =  h^{2q_a} 
 \frac{\prod_{i=1}^{q_a-1}(1-\omega^i)^{2q_a-3i} (1-\omega^{-i})^{2q_a-3i}}
 {\prod_{j=1}^{q_a}(1-\omega^{q_a+j-1})^{q_a - j +1} 
 (1-\omega^{-(q_a+j-1)})^{q_a - j +1}},
  ~~~~~~~~~ 2 \leq q_a \leq \left\lfloor \frac{r}{2} \right\rfloor. 
\end{align} 
Then $\mathbb{R}^4/\mathbb{Z}_r$ instanton partition function is 
\be
 Z^{\mathbb{R}^4/\mathbb{Z}_r}_{SU(2)} 
 = \sum_{q_a=0}^{\lfloor \frac{r}{2} \rfloor}
 \sum_{k_0=0}^{\infty} \left( {\Lambda'} \right)^{k_0+\frac{q_a(r-q_a)}{r}} 
 Z^{(r)}_{k_0 + \frac{q_a(r-q_a)}{r}}. 
\ee

\begin{figure}[h]
\begin{center}
  \includegraphics[height=2cm]{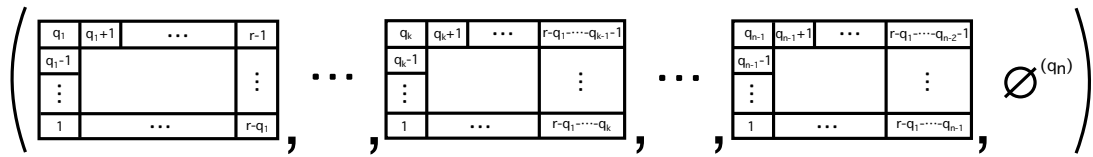}
\end{center}
\caption{Example of $n$-tuple Young diagram in the case of $k_0 = 0$, 
 $SU(n)$ and $r$. 
The total number of boxes is equal to $d$. 
\label{young2}
}
\end{figure}

In the case of $SU(n)$ and general $r$, we can start with (\ref{Nekrasov:5dim}).  From the condition (\ref{cond:chern}), we find 
\begin{align}
 & \delta k_{\ell} = \biggl(0, (n-1), \cdots, \underset{{\ell}=q_{n-1}}{q_{n-1}(n-1)}, 
 q_{n-1}(n-1)+(n-2), \cdots, \underset{{\ell}=q_{n-2}}{q_{n-1} + q_{n-2}(n-2)}, \cr
 &~~~~~
 \cdots, \underset{{\ell}=q_{n-(j-1)}}{\sum_{i=1}^{j-2} q_{n-i} + q_{n-j+1}(n-j+1)}, 
 \sum_{i=1}^{j-2} q_{n-i} + q_{n-j+1}(n-j+1) + (n-j), 
 \cr
 &~~~~~ \cdots, 
 \underset{{\ell}=q_{n-j}}{\sum_{i=1}^{j-1}q_{n-i} + (n-j)q_{n-j}}, \cdots, 
 \underset{{\ell}=q_1}{\sum_{i=1}^{n-1} q_{n-i}}, \cdots, 
 \underset{{\ell}=r - \sum_{i=1}^{n-1} q_i}{\sum_{i=1}^{n-1} q_{n-i}}, 
 \sum_{i=1}^{n-1} q_{n-i} -1, \cdots, 1
 \biggl). 
\end{align}
Here we have set 
\be
 r - \sum_{i=1}^{n-1} q_i \geq
 q_1 \geq q_2 \geq \cdots \geq q_{n-1} \geq 0 ~~~({\rm mod}~ r), 
\label{cond:charge}
\ee
All other cases on these inequalities
 can be obtained by the appropriate interchange of the numbers $q_1, 
 \cdots, q_{n-1}$. 
So we do not lose any generality.
The total number of boxes is $k_0 r + d$. 
Here
\be
d = \sum_{\alpha=1}^{n-1} {q_{\alpha}} 
\left(r - \sum_{\alpha' = 1}^{\alpha} q_{\alpha'}\right). 
\ee
Let us define
\be
 Z^{(r)}_{k_0 + \frac{d}{r}} 
 := \lim_{h \to 0} \Xi_{q_{\alpha}} 
 \sum_{\substack{|\vec{Y}|\\=rk_0+d}} 
 \widetilde{\mathcal{A}}_{\vec{Y}}^{(q_{\alpha})}. 
 \label{Nek1:ALE(r,n)}
\ee
Here $(q_{\alpha}) = (q_1, \cdots, q_{n-1}, q_n),~ 
q_n = -q_1 - \cdots - q_{n-1}$ is $\mathbb{Z}_r$-charge assigned 
 to each Young diagram in $\vec{Y}$. 
Computing the case of Fig. \ref{young2}, we obtain the coefficient 
\be
 \Xi_{q_{\alpha}} = h^{2\sum_{\alpha=1}^{n-1} \alpha q_{\alpha}} 
 \xi_{q_{\alpha}}(\omega) \xi_{q_{\alpha}}(\omega^{-1}),
\ee
\begin{align}
 \xi_{q_{\alpha}}(\omega)
 =& \prod_{a < b}^{n-1} \prod_{j=1}^{q_{b}-1} 
 \frac{(1 - \omega^j)^{2(q_b -j)}}
 {(1-\omega^{\sum_{i=a}^{b-1}q_{i}+j})^{q_b-j}
 (1-\omega^{\sum_{i=a}^{b-1}q_{i}+j})^{q_b-j}}
 \frac{1}{(1 - \omega^{\sum_{i=a}^{b-1}q_{i}})^{q_b}} \cr 
 &\times \prod_{a=1}^{n-1} 
 \prod_{j=1}^{q_a-1} \frac{(1-\omega^j)^{q_a-j}
 (1-\omega^{\sum_{i=a+1}^{n-1}q_i+j})^{q_a-j}}
 {(1-\omega^{\sum_{i=1}^{a-1} q_i+j})^j
  (1 - \omega^{\sum_{i=1}^{n-1}q_i+j})^{q_a-j}} 
 \frac{1}{( 1- \omega^{\sum_{i=1}^{a} q_i})^{q_a}} \cr
 &\times\prod_{a=1}^{n-1}  \prod_{j=1}^{q_{a+1} + \cdots q_{n-1} }
 \frac{(1-\omega^j)^{q_a}}{(1-\omega^{\sum_{i=1}^{a} q_i + j})^{q_a}} \cr
 &\times \prod_{a+2\leq b}^{n-1} \prod_{j=1}^{q_b-1}
 \frac{(1-\omega^j)^j (1-\omega^{\sum_{i=a+1}^{b-1} q_i +j})^{q_b-j}}
 {(1-\omega^{q_a-q_b+j})^j (1-\omega^{\sum_{i=k}^{l-1}q_i-j})^j} 
 \prod_{j=0}^{q_{a+1}+\cdots+q_{b-1}-q_b} 
 \frac{(1-\omega^{q_b+j})^{q_b}}{(1-\omega^{q_a+j})^{q_b}}.
\end{align}
Then the $\mathbb{R}^4/\mathbb{Z}_r$ instanton partition function is given by 
\be
Z^{\mathbb{R}^4/\mathbb{Z}_r}_{SU(n)} 
 = \sum_{q_{\alpha}}
\sum_{k=0}^{\infty} (\Lambda')^{k+\frac{d}{r}}
 Z^{(r)}_{k+ \frac{d}{r}}. 
\label{Nek2:ALE(r,n)}
\ee

To summarize, the method of obtaining the $SU(n)$ instanton partition function 
 on $\mathbb{R}^4/\mathbb{Z}_r$ from the five dimensional instanton partition function
 by the limiting procedure (\ref{limit:gauge}) has been established. 
The ALE instanton partition functions are systematically obtained 
 by using (\ref{Nek1:ALE(r,n)})-(\ref{Nek2:ALE(r,n)}).


\section{Note added} 


References we became aware after submission include \cite{EPSS,PS,CMM}.

\section*{Acknowledgements}
We  thank Hiroaki Kanno, Katsushi Ito, Sanefumi Moriyama, Kazuhiro Sakai, 
 Hirotaka Irie and Toshio Nakatsu for valuable discussions. 
The authors' research is supported in part by the Grant-in-Aid 
for Scientific Research from the Ministry of Education, Science and Culture, 
Japan (23540316).
Support from JSPS/RFBR bilateral collaboration 
``Progress in the synthesis of integrabilities arising from gauge-string duality''
(FY2012-2013: 12-02-92108-Yaf-a) is gratefully appreciated.



\end{document}